\documentclass[amsmath,trackchanges,twocolumn]{aastex702}

\usepackage{graphicx}
\usepackage{longtable}
\usepackage{booktabs}
\usepackage{booktabs}
\usepackage{microtype}
\usepackage[noabbrev,capitalize]{cleveref}
\usepackage{silence}
\usepackage{subfig}
\usepackage{soul}
\usepackage{glossaries-extra}

\newcommand{\gsfcAffiliationString}{NASA Goddard Space Flight Center, Greenbelt, MD 20771, USA}
\newcommand{\usraAffiliationString}{Universities Space Research Association, Columbia, MD 21046, USA}
\newcommand{\umdAffiliationString}{Department of Physics, University of Maryland, College Park, MD 20742, USA}
\newcommand{\calAffiliationString}{Space Sciences Laboratory, University of California, Berkeley, Berkeley, CA 94720, USA}

\newacronym[
    first={\textit{Transiting Exoplanet Survey Satellite}
    (\textit{TESS}, \citealp{ricker2015transiting})}
]{TESS}{\textit{TESS}}{\textit{Transiting Exoplanet Survey Satellite}}

\begin{document}
    \title{Heartbeat binary catalog identified via convolutional neural networks}

\author[0009-0004-7491-2042]{Georgia Kalapotharakou}
\affiliation{\umdAffiliationString}
\email{gkalapot@terpmail.umd.edu}

\author[0000-0001-8472-2219]{Greg Olmschenk}
\affiliation{\gsfcAffiliationString}
\affiliation{\usraAffiliationString}
\email{gregory.olmschenk@nasa.gov}

\author{Siddhant Solanki}
\affiliation{\gsfcAffiliationString}
\email{-}

\author{Jeremy Schnittman}
\affiliation{\gsfcAffiliationString}
\email{-}

\author{Agnieszka M. Cieplak}
\affiliation{\calAffiliationString}
\email{-}

\begin{abstract}
We present a catalog of 688 heartbeat binary candidates identified by a convolutional neural network (CNN) in Transiting Exoplanet Survey Satellite (TESS) data. We modeled the phase-folded light curves using parallel-tempered Markov Chain Monte Carlo (MCMC) to estimate stellar masses, orbital eccentricity, argument of periastron, and other stellar and orbital parameters. Systems with a median posterior eccentricity of at least 0.2 were retained for population analysis. We construct the period-eccentricity distribution of the final population and find that the upper eccentricity envelope increases toward longer orbital periods. Using stellar properties derived from the median posterior parameters of the MCMC fits, we also examine the temperature, spectral-type, and radius distributions of the primary and secondary stars. The inferred primary-star temperatures span a broad range, while a large fraction of the secondary components have inferred temperatures corresponding to M-type stars and radii typically less than one solar radius. This catalog substantially expands the population of heartbeat binary candidates identified in TESS data and provides a large sample for studying the stellar properties and evolution of eccentric binary systems. 
\end{abstract}
    \section{Introduction}
\label{sec:introduction}

Heartbeat binaries are detached binary star systems with eccentric orbits that exhibit characteristic photometric variability near periastron, when the stars reach their closest separation. The increased tidal forces at periastron temporarily distort the stars, producing changes in their observed brightness. Depending on the orbital geometry, these variations can produce a distinctive feature in the light curve resembling an electrocardiogram signal, from which heartbeat binaries derive their name \citep{thompson2012class}.

The shape and amplitude of the heartbeat signal depend on several properties of the system, including orbital eccentricity, orbital orientation, stellar radii, and the relative brightness of the two stars \citep{thompson2012class}. Some heartbeat systems also exhibit eclipses when one star passes in front of the other along the observer's line of sight. The characteristic heartbeat feature is primarily associated with the changing tidal distortion of the stars near periastron. In addition, dynamical tides can excite stellar oscillations, known as tidally excited oscillations (TEOs), which often occur at integer multiples of the orbital frequency and introduce additional variability into the light curve \citep{fuller2017heartbeat}.

Heartbeat binaries provide an opportunity to study the effects of tidal interactions on the orbital evolution of eccentric binary systems. Over time, tidal dissipation can reduce orbital eccentricity and eventually lead to circularization. The characteristic photometric variability of heartbeat binaries provides information about their orbital properties and tidal interactions, making these systems valuable for investigating binary evolution. In systems that also exhibit eclipses, the additional photometric features can place stronger constraints on orbital inclination, stellar radii, and other system parameters.

The study of heartbeat binaries advanced significantly with the development of high-precision photometric surveys. In particular, observations from the Kepler space telescope enabled the discovery of numerous heartbeat binaries and detailed investigations of their photometric variability \citep{thompson2012class}. Ground-based surveys have also contributed substantially to the identification of these systems. Using observations from the Optical Gravitational Lensing Experiment (OGLE), \citet{wrona2022ogle} identified 991 heartbeat binary candidates in the Galactic bulge and Magellanic Clouds, significantly expanding the known population. However, the limited sky coverage of the original Kepler mission and the atmospheric effects that can limit the photometric precision of ground-based surveys motivate additional searches for heartbeat binaries using TESS, which combines high-precision space-based photometry with broad sky coverage.

\gls{TESS} provides an opportunity to search for heartbeat binaries across a much larger fraction of the sky \citep{ricker2015transiting}. Although primarily designed to detect transiting exoplanets, its high-precision photometry also makes it a valuable tool for studying variable stars and binary systems. Many targets are observed in multiple sectors, sometimes separated by hundreds of days, providing extended observational baselines that can improve the determination of orbital periods and the characterization of periodic variability. Previous studies have already demonstrated the potential of \gls{TESS} for identifying heartbeat binaries. In particular, \citet{solanki2025short} identified 240 short-period binary systems in \gls{TESS} data, 180 of which were classified as heartbeat binaries.

\gls{TESS} observes millions of stars across the sky, making a systematic manual search for heartbeat binaries impractical. For example, \citet{kunimoto2021quick} reported light curves for more than 9 million unique stars brighter than a TESS magnitude of $T=13.5$ during the first year of the extended mission. In this work, we conduct a large-scale search for heartbeat binaries using a convolutional neural network (CNN) applied to approximately 4.4 million TESS-SPOC light curves. The CNN was trained to recognize heartbeat-like variability and used to identify the 5,000 highest-ranked unique targets for manual inspection. Following visual vetting, the selected candidates were analyzed using a period-search algorithm and a parallel-tempered Markov Chain Monte Carlo (MCMC) approach to model their photometric variability and estimate their stellar and orbital parameters. We present a catalog of heartbeat binary candidates and investigate their orbital period, eccentricity, and inferred stellar properties to characterize the identified population.
    \section{Observational Data}
\label{sec:observational-data}

Heartbeat binaries are identified through their characteristic, repeating changes in brightness, making stellar light curves the primary observational data for this search \citep{thompson2012class,shporer2016radial,kirk2016kepler}. \gls{TESS} observes the sky using four wide-field cameras, each covering approximately $24^\circ \times 24^\circ$, providing a combined field of view of $24^\circ \times 96^\circ$ \citep{ricker2015transiting}. The sky is observed in successive regions known as sectors, each typically monitored for approximately 27 days. The original two-year mission consisted of 26 observing sectors, covering the southern and northern ecliptic hemispheres, while subsequent extended missions have continued observing additional and previously observed regions of the sky.

Although each sector is observed for approximately 27 days, observations are not entirely continuous. During the observing cycles considered in this work, data collection was briefly interrupted near perigee, approximately every 13.7 days, to transmit the collected data to Earth \citep{ricker2015transiting}. In addition, neighboring sectors overlap, particularly near the ecliptic poles, allowing some targets to be observed over multiple consecutive sectors. Other targets may be revisited after gaps of months or years, depending on the spacecraft pointing and observing schedule. 

During Sectors 27–55, TESS full-frame images (FFIs) were obtained at a 10-minute cadence. The Science Processing Operations Center (SPOC) at NASA Ames Research Center processes TESS observations using a pipeline developed from the \textit{Kepler} data-processing pipeline \citep{jenkins2016tess}. The SPOC pipeline calibrates the image data and produces photometric data products, including light curves with corrections for instrumental systematics. The TESS-SPOC data product extends this processing to targets extracted from the FFIs. We used the TESS-SPOC light-curve products generated from the FFIs, which provide flux measurements as a function of time.

We analyzed SPOC light curves from \gls{TESS} Sectors 27-55. The resulting dataset contained 4,398,877 \gls{TESS}-SPOC sector light curves, corresponding to 2,335,056 unique TESS Input Catalog (TIC) targets. Because some targets were observed in multiple sectors, individual stars could contribute more than one light curve to the dataset. In total, 922,103 TIC targets appeared more than once, while 1,412,953 targets appeared only once. These light curves were used for the neural network search after being preprocessed to a uniform length and normalized using a modified z-score based on the median and median absolute deviation. Our study extends the work of \citet{solanki2025short} by conducting a dedicated search for heartbeat binaries using uniformly processed \gls{TESS}-SPOC light curves and examining a substantially larger sample of candidates.

The data were downloaded as FITS files. From each file, we used the observation time, corrected flux, and flux uncertainty. The time values are given in BTJD, which is defined as BJD $- 2457000.0$, where BJD is the Barycentric Julian Date. 

We used the Presearch Data Conditioning Simple Aperture Photometry (PDCSAP) light curves provided by the TESS-SPOC pipeline. These light curves are derived from the Simple Aperture Photometry (SAP) measurements after applying corrections intended to reduce instrumental systematic effects \citep{jenkins2016tess}. The flux values were normalized by dividing by the median flux of that sector, and the flux uncertainties were normalized by dividing by the absolute value of the same median flux, before the period search and modeling. This sector-by-sector normalization was performed after candidate selection by the CNN described below for the period search, phase folding, and MCMC modeling, rather than for the neural network search.

\begin{figure*}[t]
    \centering

    \begin{minipage}{0.48\textwidth}
        \centering
        \includegraphics[width=\linewidth]{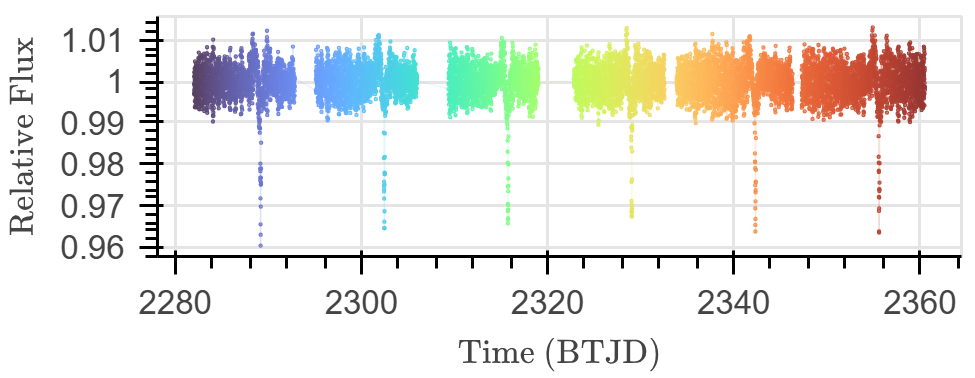}
        \par (a)
    \end{minipage}

    \vspace{0.3cm}

    \begin{minipage}{0.48\textwidth}
        \centering
        \includegraphics[width=\linewidth]{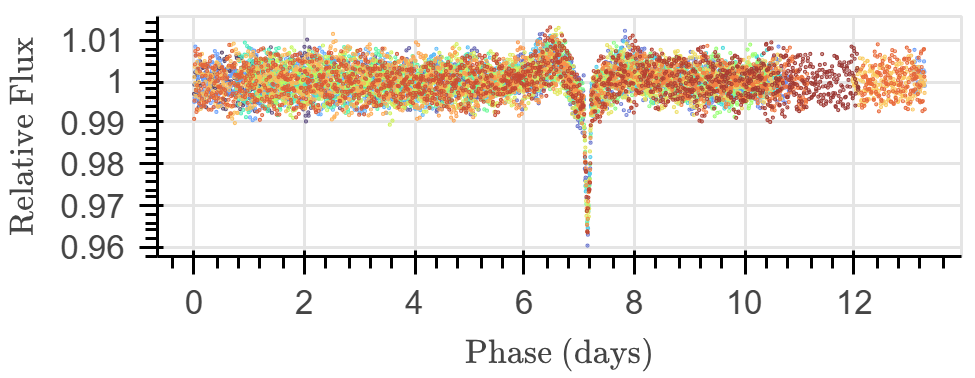}
        \par (b)
    \end{minipage}
    \hfill
    \begin{minipage}{0.48\textwidth}
        \centering
        \includegraphics[width=\linewidth]{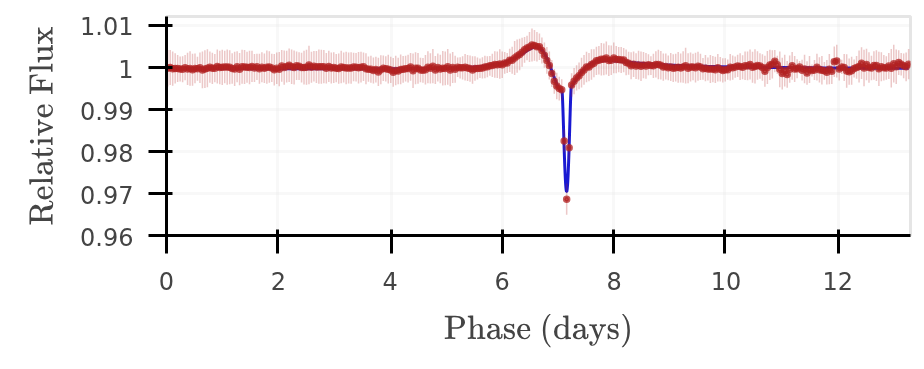}
        \par (c)
    \end{minipage}

    \caption{\gls{TESS} observations of TIC 302795956.
    (a) Corrected light curves from Sectors 36, 37, and 38 stitched together, with data points colored according to observation time.
    (b) Light curve folded by its best period, retaining the original observation-time colors.
    (c) Folded light curve with 300 bins along with their corresponding errors shown in red, and the best-fit model obtained from the MCMC analysis shown in blue.}
    \label{fig:heartbeat-example}
\end{figure*}

For each target, we combined the sectors in which the TIC was observed after normalizing the data from each sector. We then determined the best period and folded the combined light curve on that period. The folded light curves were binned using 300 bins for multi-sector light curves and 150 bins for single-sector light curves. The final folded and binned light curves were used for heartbeat-binary modeling. Figure~\ref{fig:heartbeat-example} illustrates this process for TIC 302795956, showing the stitched light curve, the phase-folded observations, and the resulting best-fit MCMC model.

    \section{Methodology}
\label{sec:methodology}

Our analysis consists of two main stages: the identification of heartbeat-binary candidates and the inference of their stellar and orbital properties. We train a convolutional neural network (CNN) specifically for this study using the synthetic heartbeat and non-heartbeat signals generated by \citet{solanki2025short}, which are injected into observed TESS-SPOC light curves from Sectors 27--55. The trained CNN is applied to approximately 4.4 million TESS-SPOC light curves, corresponding to approximately 2.3 million unique targets, and assigns each light curve an uncalibrated classification score between 0 and 1. We rank the targets according to their highest CNN score across the available sectors and select the top 5,000 unique targets for further analysis.

For each selected target, we combine observations across available sectors, estimate its orbital period using the Box Least Squares (BLS) algorithm, and visually inspect its unfolded and phase-folded light curves to assess the presence of recurring heartbeat-like variability. The 1,100 targets pass this stage are modeled using a physical binary-star model with parallel-tempered Markov Chain Monte Carlo (MCMC) to infer stellar and orbital parameters, keeping the estimated orbital periods fixed. The resulting fits are visually assessed, supported by quantitative measures of fit quality, including a metric developed in this work to identify coherent discrepancies between the modeled and observed light curves. This process yields 732 systems with acceptable fits. Finally, we retain systems with median posterior eccentricities of $e \geq 0.2$, producing a final sample of 688 heartbeat-binary candidates. The inferred stellar and orbital properties of these systems form the basis for the population analysis presented below.

\subsection{CNN Training and Validation}
We use the NN architecture defined in \citet{Olmschenk_2021}. That is, the \texttt{Hadryss} convolutional neural network (CNN) architecture implemented in the \texttt{qusi} framework (\url{https://github.com/golmschenk/qusi}). For details, see \citet{Olmschenk_2021}, but briefly, the network consists of 9 one-dimensional convolutional layers and 3 dense layers. It takes as input a light curve containing 3,500 flux measurements. The architecture uses convolutional layers to identify features in the light curves across progressively larger scales, and the dense layers at the end to encapsulate globally distilled features. A final sigmoid activation function produces a classification score between 0 and 1. The CNN was trained to recognize heartbeat-like variability using labeled examples constructed from observed \gls{TESS}-SPOC light curves and synthetic signal injections, again following the general approach described by \citet{Olmschenk_2021}.

The synthetic signals used in this work were obtained from the training dataset developed by \citet{solanki2025short}. The synthetic heartbeat light curves were generated using their implementation of the eccentric BEaming Ellipsoidal Reflection (eBEER) model, which accounts for photometric variations arising from tidal distortion, Doppler beaming, reflection effects, and eclipses, including limb-darkening effects. The model produces a variety of heartbeat signal morphologies depending on the stellar and orbital parameters of the binary system.

The dataset consisted of three components: 200,000 synthetic heartbeat signals used for positive injections, 200,000 synthetic non-heartbeat signals used for negative injections, and approximately 4 million unmodified observed light curves labeled as negative. The unmodified observed light curves were not independently screened for heartbeat-like variability before being assigned negative labels. Instead, this labeling choice was based on the assumption that heartbeat binaries constitute a small fraction of the observed population. However, due to the statistically small proportion of such label contamination, this is not expected to present an issue in training. Using observed light curves as the background for both positive and negative injections allowed the network to encounter the synthetic signals in the presence of realistic photometric scatter, intrinsic stellar variability, and residual observational effects. No additional stochastic scatter was added to the synthetic model light curves themselves, as the real base light curves being injected into already contain real noise.

The light curves used for CNN training and subsequent inference underwent a common preprocessing procedure. Nonfinite flux measurements, including NaN and infinite values, were removed, and light curves containing no valid measurements were excluded. Each light curve was then standardized to a uniform length of 3,500 data points. Shorter light curves were extended by repeating their data, whereas longer light curves were truncated. Finally, the flux values were normalized using a modified z-score based on the median and median absolute deviation
\citep{iglewicz1993volume}.

Sector 38 was excluded from the observed light curves used to construct the CNN training dataset because it exhibited increased photometric scatter during initial inspection, consistent with documented scattered-light contamination during the sector \citep{fausnaugh2021sector38}. This exclusion did not apply to subsequent CNN inference. The synthetic signal collections were shuffled using a fixed random seed. Each of the three underlying data collections was then divided separately into training, validation, and test subsets containing 80\%, 10\%, and 10\% of the data, respectively. For validation, examples were sampled in approximately equal proportions from the three categories: synthetic heartbeat injections, synthetic non-heartbeat injections, and unmodified observed light curves. Consequently, approximately one-third of the evaluated examples were labeled positive and two-thirds negative, despite the substantially different sizes of the underlying categories.

The CNN was trained using a binary cross-entropy loss function and the AdamW optimizer. Training was performed using a batch size of 1,000, with 100 training steps and 10 validation steps per cycle. Its classification performance was evaluated using binary accuracy and the area under the receiver operating characteristic curve (AUROC). The model achieved a validation accuracy of approximately 85.3\% and a validation AUROC of 0.909. As a baseline comparison, with 2/3 of the validation data consisting of negative samples, a classifier that always predicted the negative class would achieve an accuracy of approximately 66.7\%. The AUROC of 0.909 indicates that the network assigns a higher score to a randomly selected positive example than to a randomly selected negative example approximately 90.9\% of the time. This discrimination capability is particularly relevant to the use of the CNN as a candidate-ranking tool. 

However, these metrics characterize its performance on the constructed validation sample and do not directly establish the completeness of the resulting heartbeat-binary catalog.

\subsection{Candidate Selection}
\begin{figure}
    \centering
    \includegraphics[width=0.9\linewidth]{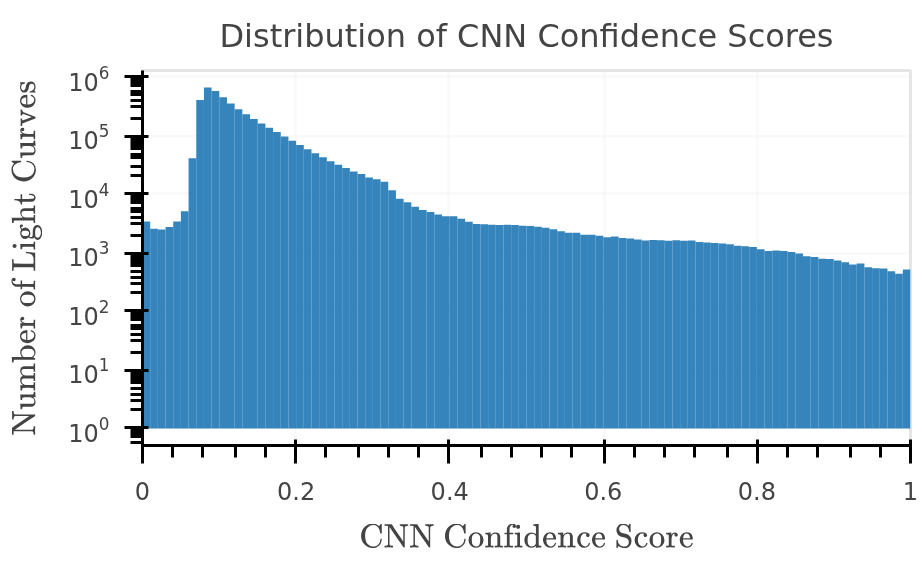}
    \caption{Distribution of CNN (uncalibrated) confidence scores assigned to 4,398,877 TESS-SPOC sector light curves. The vertical axis shows the number of light curves in each confidence-score bin on a logarithmic scale. Most light curves receive relatively low scores, with the distribution peaking near 0.1, while a smaller number extend toward higher confidence values.}
    \label{fig:cnn-confidence-scores}
\end{figure}
\begin{figure}
    \centering
    \includegraphics[width=0.9\linewidth]{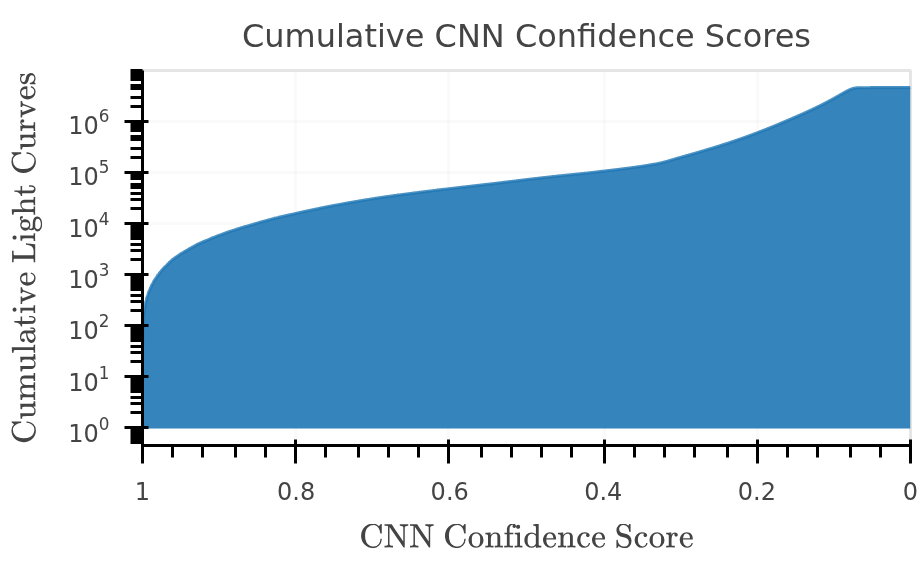}
    \caption{Reverse cumulative distribution of CNN (uncalibrated) confidence scores for 4,398,877 TESS-SPOC sector light curves. The vertical axis shows the number of light curves with confidence scores greater than or equal to the corresponding value on the horizontal axis, displayed on a logarithmic scale. The CNN confidence scores were subsequently used to select the 5,000 highest-ranked unique TICs for period estimation and visual inspection.}
    \label{fig:cnn-confidence-cumulative-scores}
\end{figure}

After training, the neural network model was used for inference on the \gls{TESS}-SPOC light curves. For inference, each light curve was passed through the trained model, which returned a confidence score for how likely the light curve was to contain a heartbeat binary signal. The trained CNN was applied to approximately 4.4 million TESS-SPOC sector light curves from Sectors 27--55, producing a confidence score between 0 and 1 for each light curve. Both because it is not the goal of this work, and because real world occurrence rates are unknown, we do not calibrate the CNN confidences. That is, these confidences only provide an ordered ranking, and do not correspond to real world detection rates. Figure~\ref{fig:cnn-confidence-scores} shows the distribution of these scores, while Figure~\ref{fig:cnn-confidence-cumulative-scores} shows the corresponding reverse cumulative distribution, representing the number of light curves with confidence scores greater than or equal to a given threshold. The majority of light curves received relatively low confidence scores, with the distribution peaking near 0.1 and decreasing toward higher scores. A smaller population extends to scores approaching 1, indicating that a limited fraction of the observed light curves exhibit features that the CNN associates more strongly with synthetic heartbeat signals. The continuous distribution of scores, without a clearly separated high-confidence population, suggests that the network does not provide an obvious confidence threshold that uniquely distinguishes heartbeat binaries from other sources of variability.

We selected the top 5,000 unique TIC IDs ranked by CNN confidence score, ranking each target by its highest score across all observed sectors. This reduced the initial set of 4,398,877 sector light-curve evaluations to 5,000 unique targets for further examination, corresponding to a reduction in the number of entries requiring consideration by a factor of approximately 880. The selection of 5,000 unique targets was therefore a practical compromise between retaining a substantial set of high-ranking candidates and maintaining a manageable human workload. This cutoff was not derived from a statistically determined separation between heartbeat and non-heartbeat populations, nor does it guarantee that all detectable heartbeat systems were included. Consequently, the CNN was used primarily as a candidate-ranking and filtering tool, with the final identification of heartbeat candidates requiring additional visual examination and physical parameter fitting using MCMC. Manual vetting was necessary because a high CNN confidence score did not always correspond to a genuine heartbeat signal. Approximately 1,100 of the 5,000 visually inspected candidates (22\%) were retained for physical modeling, while the remaining 78\% were excluded during manual vetting.

\begin{figure}
    \centering
    \includegraphics[width=0.9\linewidth]{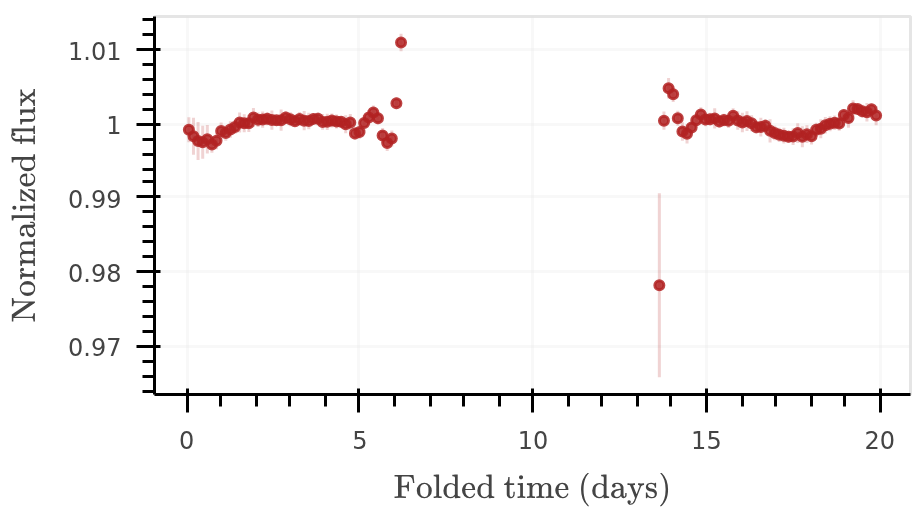}
    \caption{Folded \gls{TESS} light curve of TIC 181923581 showing a substantial gap in phase coverage. This candidate was rejected during manual vetting because the incomplete coverage prevented a reliable assessment of the heartbeat signal.}
    \label{fig:coverage-gap}
\end{figure}

\begin{figure}
    \centering
    \includegraphics[width=0.9\linewidth]{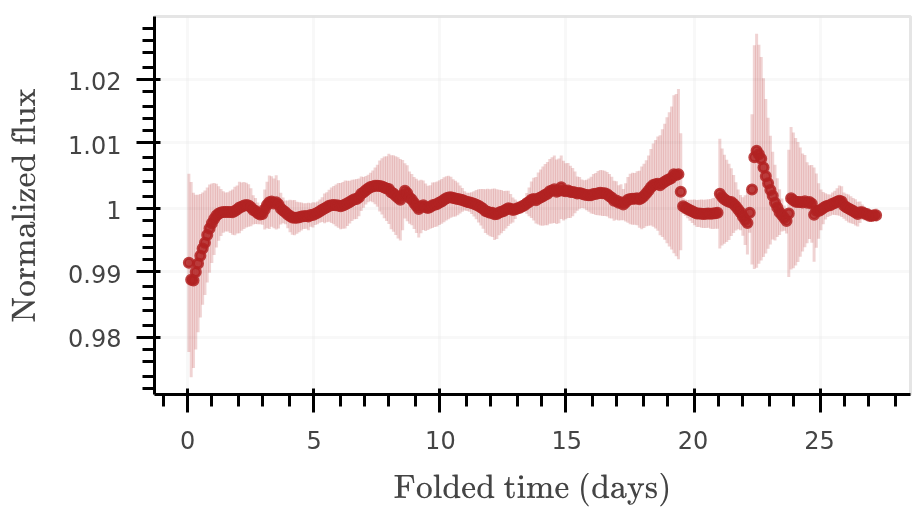}
    \caption{Folded \gls{TESS} light curve of TIC 229641091 showing a signal that does not exhibit coherent heartbeat-like variability. This candidate was rejected during manual vetting.}
    \label{fig:not-coherent}
\end{figure}

During manual vetting, we excluded light curves with substantial gaps in phase coverage as shown in Figure~\ref{fig:coverage-gap}, light curves that did not resemble heartbeat signals as shown in Figure~\ref{fig:not-coherent}, and light curves that were poorly phase-folded. We also excluded candidates that exhibited only a single heartbeat-like event during the available observations, since the lack of repeated events prevented us from reliably determining their orbital periods.

\begin{figure}
    \centering
    \includegraphics[width=0.9\linewidth]{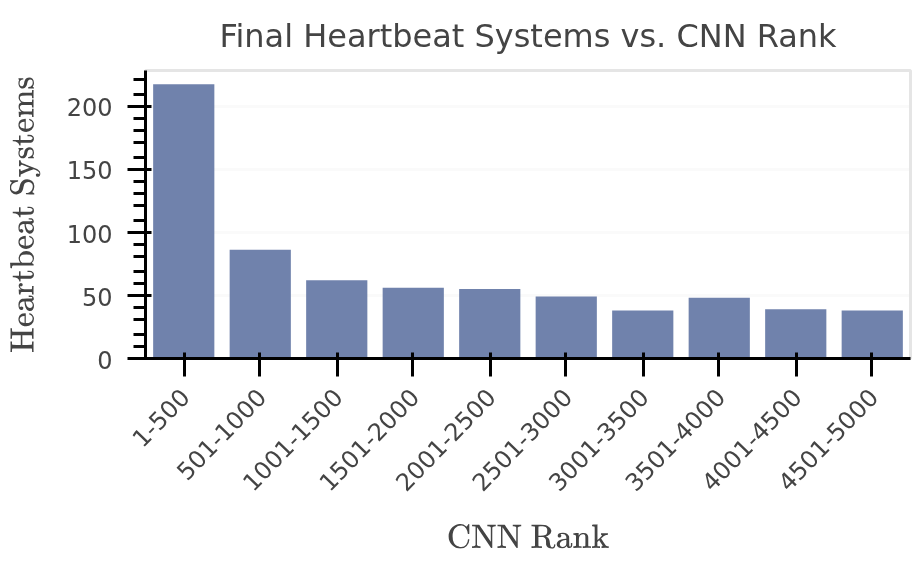}
    \caption{Distribution of the final heartbeat binary candidates as a function of their original CNN rank. The 5,000 highest-ranked unique TICs were divided into ten consecutive groups of 500, with lower rank numbers corresponding to higher CNN confidence scores. The number of candidates retained in the final sample is shown for each group. The highest-ranked 500 light curves contain the largest number of retained heartbeat systems, although candidates are identified throughout the full ranking range.}
    \label{fig:hb-rank}
\end{figure}

Figure~\ref{fig:hb-rank} shows the distribution of the final heartbeat binary candidates across the original CNN rankings. The largest number of retained systems was found among the 500 highest-scoring light curves, with the number generally decreasing toward lower-ranked groups. Nevertheless, heartbeat candidates were identified throughout the top 5,000, indicating that limiting the search to only the highest-ranked few hundred light curves would have excluded a substantial number of candidates. Extending the search beyond the top 5,000 would likely yield additional candidates, although at the cost of increasing manual vetting effort per newly identified heartbeat system.

The purpose of the neural network was not to create a final catalog of heartbeat binaries by itself, but to significantly reduce the number of light curves that needed to be inspected manually. The original dataset contained over four million light curves, and a classifier helped reduce the search to a much smaller set of high confidence candidates.

\subsection{Period Search and Folding}
For each candidate identified during visual inspection, we retrieved all available \gls{TESS} light curves from Sectors 27--55. For each sector in which the target was observed, we extracted the observation times, corrected flux measurements, and corresponding flux uncertainties, represented by the \texttt{TIME}, \texttt{PDCSAP\_FLUX}, and \texttt{PDCSAP\_FLUX\_ERR} columns of the SPOC FFI light curve data.

Measurements containing nonfinite values of time, flux, or flux uncertainty were removed, along with measurements having zero or negative flux uncertainties. Each sector's light curve was normalized independently by dividing the flux measurements by the median flux of that sector, while the corresponding uncertainties were scaled by the absolute value of the same median. For each target, the normalized light curves from all available sectors were then combined to construct a single light curve.

We estimated the orbital period of each candidate using the Box Least Squares (BLS) algorithm \citep{kovacs2002box}, which searches for periodic, box-shaped features in photometric time series. For each candidate, we combined all available TESS light curves from sectors 27–55, using the PDCSAP flux measurements. Each sector's flux and associated uncertainties were normalized by the median flux of that sector before combining the observations. Because heartbeat binary light curves can exhibit both increases and decreases in brightness, we performed two independent BLS searches for each candidate: one using the original normalized flux and another using the inverted flux, obtained by multiplying the flux measurements by $-1$. The inverted search allows BLS to identify periodic brightening features as decreases in the inverted flux. Both searches evaluated 10,000 uniformly spaced trial periods between 1 and 30 days, using a box duration of 0.1 days. The period corresponding to the maximum BLS power was selected from each search, yielding two candidate periods.

To determine which period produced a more coherent folded light curve, we phase-folded the original observations separately using both candidate periods and compared the scatter of the resulting folds. The observations were divided into equally spaced phase bins, with 300 bins used for targets observed in multiple sectors and 150 bins for targets observed only in a single sector. For each bin containing at least three measurements, we calculated the mean normalized flux and fitted a linear trend to the individual flux measurements as a function of orbital phase. We then calculated the standard deviation of the residuals about this linear fit to characterize the local scatter. The linear fit is used to help prevent true variation from being considered scatter. Bins containing fewer than three measurements were excluded because at least three points were required to estimate the residual scatter after fitting a line. For each candidate period, we calculated the 95th percentile (P95) of these residual standard deviations across all valid phase bins, where a lower P95 indicates less scatter. Because BLS is designed to identify transit-like features rather than the full range of heartbeat morphologies, the highest-power BLS period does not necessarily correspond to the true orbital period. Furthermore, small differences in scatter between two folds do not necessarily indicate that one period is more accurate. Therefore, we introduced a heuristic threshold of 1.2, selecting the inverted BLS period only when the P95 of the original fold was at least 1.2 times that of the inverted-period fold. Otherwise, the original BLS period was retained.

After selecting the preferred period, we retained the corresponding phase-folded and binned light curve for subsequent visual inspection and physical modeling. The mean normalized flux of each valid bin was used as the binned flux value, while the standard deviation of the residuals from the linear fit was used as its uncertainty in the MCMC likelihood. A minimum uncertainty of $10^{-4}$ was imposed, and the noise-rescaling parameter was fixed at zero, so no additional noise rescaling was applied. The resulting binned light curves were then used as inputs to the MCMC modeling.

Given the large number of candidates and the subsequent visual assessment of the phase-folded light curves, we adopted a practical and consistent period-estimation procedure rather than attempting an exhaustive optimization for each system. In particular, we used a fixed BLS duration of 0.1 days, although heartbeat features can vary in duration between systems. Additionally, the 1.2 threshold used to select between the original and inverted BLS periods was chosen based on preliminary visual comparisons, with the aim of avoiding period selections driven by small differences in the scatter of the folded light curves. Although these choices were not systematically optimized, they provide a consistent procedure for obtaining initial period estimates that are subsequently assessed through visual inspection. Evaluating a wider range of BLS durations, testing different selection thresholds, and comparing additional candidate periods could potentially improve period recovery for some systems, but establishing whether these changes consistently improve accuracy would require further testing and validation. Our approach therefore represents a practical period-estimation method rather than a fully optimized one. 

\subsection{Heartbeat Binary Modeling}

Following the period search and phase-folding procedure,
each candidate was modeled using the \texttt{binara}
heartbeat binary modeling code \citep{solanki2025short}.

The modeling was performed using a parallel-tempered MCMC
method. For each proposed set of model parameters,
\texttt{binara} generated a synthetic light curve and
compared it with the observed phase-folded light curve
using a chi-squared-based log-likelihood function. 
For each candidate, the MCMC was run for $10^6$ iterations.
The evolution of the median log-likelihood and sampled
parameter distributions was examined, with both remaining
approximately stable over most of the run.

The orbital and stellar parameters included the logarithms of the primary and secondary masses (in solar masses), the logarithm of the orbital period (in days), eccentricity, orbital inclination, argument of periastron, and a reference epoch. The orbital period was held fixed at the value determined by the period search during MCMC fitting. Additional parameters included an overall flux-scaling factor, logarithmic radius-scaling factors for both stellar components, and the coefficients $\mu_1$, $\tau_1$, $\mu_2$, $\tau_2$, $\alpha_{\mathrm{ref},1}$, $\alpha_{\mathrm{ref},2}$, $\alpha_{\mathrm{beam},1}$, $\alpha_{\mathrm{beam},2}$, $\alpha_{T,1}$, and $\alpha_{T,2}$. The model also included blending and flux-tuning parameters. The adopted prior distributions and parameter bounds are summarized in Table~\ref{tab:model-priors}.

\begin{table*}
\centering
\caption{Model parameters priors used in the MCMC fitting. Each parameter is either fixed or follows a gaussian or uniform distribution. Gaussian priors are truncated at the parameter bounds.}
\label{tab:model-priors}
\begin{tabular}{llll}
\toprule
Parameter & Description & Distribution & Bounds\\
\midrule
$\log M_1$ & Log mass of the primary star in solar masses &  $\mathcal{U}(-1.5,2)$ & $[-1.5,2]$\\
$\log M_2$ & Log mass of the secondary star in solar masses &  $\mathcal{U}(-1.5,2)$ & $[-1.5,2]$ \\
$\log P$ & Period & fixed & $[-2,3]$ \\
$e$ & Eccentricity & $\mathcal{U}(0,1)$ & $[0,1]$\\
$\cos i$ & Orbital inclination &  $\mathcal{U}(0,1)$ & $[0,1]$\\
$\omega$ & Argument of periastron &  $\mathcal{U}(-\pi,\pi)$ & $[-\pi,\pi]$\\
$T_0$ & Epoch of periastron passage &  $\mathcal{U}(0,2\pi)$ & $[0, 2\pi]$\\
$\log r_1$ & Radius re-scaling parameter for star 1 & $\mathcal{N}(0,1)$ & $[-5,5]$\\
$\log r_2$ & Radius re-scaling parameter for star 2 & $\mathcal{N}(0,1)$ & $[-5,5]$\\
$\mu_1$ & Limb-darkening parameter of star 1 & $\mathcal{N}(0.2,0.2)$ & $[0,1]$\\
$\tau_1$ & Gravity-darkening parameter of star 1 & $\mathcal{N}(0.4,0.2)$ & $[0.15, 0.55]$\\
$\mu_2$ & Limb-darkening parameter of star 2 & $\mathcal{N}(0.2,0.2)$ & $[0,1]$ \\
$\tau_2$ & Gravity-darkening parameter of star 2 & $\mathcal{N}(0.4,0.2)$ & $[0.15, 0.55]$\\
$\alpha_{\mathrm{ref},1}$ & Reflection coefficient of star 1 & $\mathcal{N}(1,0.2)$ & $[0,2]$ \\
$\alpha_{\mathrm{ref},2}$ & Reflection coefficient of star 2 & $\mathcal{N}(1,0.2)$ & $[0,2]$\\
$\alpha_{\mathrm{beam},1}$ & Doppler beaming coefficient of star 1 & $\mathcal{N}(0,0.1)$ & $[-0.3, 0.3]$\\
$\alpha_{\mathrm{beam},2}$ & Doppler beaming coefficient of star 2 & $\mathcal{N}(0,0.1)$ & $[-0.3, 0.3]$\\
$\alpha_{T,1}$ & Temperature re-scaling parameter for star 1 & $\mathcal{N}(0,1)$ & $[-5,5]$\\
$\alpha_{T,2}$ & Temperature re-scaling parameter for star 2 & $\mathcal{N}(0,1)$ & $[-5,5]$\\
$\delta_s$ & Sector blending parameter & $\mathcal{U}(0,1)$ & $[0,1]$\\
$f_{\mathrm{tune}}$ & Sector flux normalization &  $\mathcal{U}(0.99,1.01)$ & $[0.99, 1.01]$\\
$\ln \sigma_{\mathrm{resc}}$ & Sector noise re-scaling parameter &  fixed & $[0,0]$\\
\bottomrule
\end{tabular}
\end{table*}

\subsection{Evaluating Model Fits}
\begin{figure}
    \centering
    \includegraphics[width=0.9\linewidth]{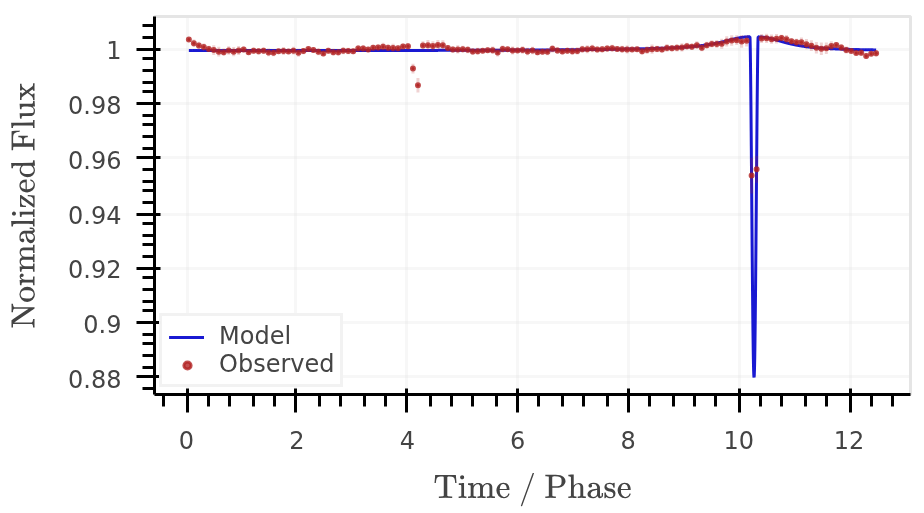}
    \caption{Folded \gls{TESS} light curve of TIC 20176241. This candidate was rejected during manual vetting because the heartbeat-binary model fails to reproduce the observed eclipse features. In particular, the model does not capture the apparent secondary eclipse and substantially overestimates the depth of the primary eclipse.}
    \label{fig:missing-dip}
\end{figure}

\begin{figure}
    \centering
    \includegraphics[width=0.9\linewidth]{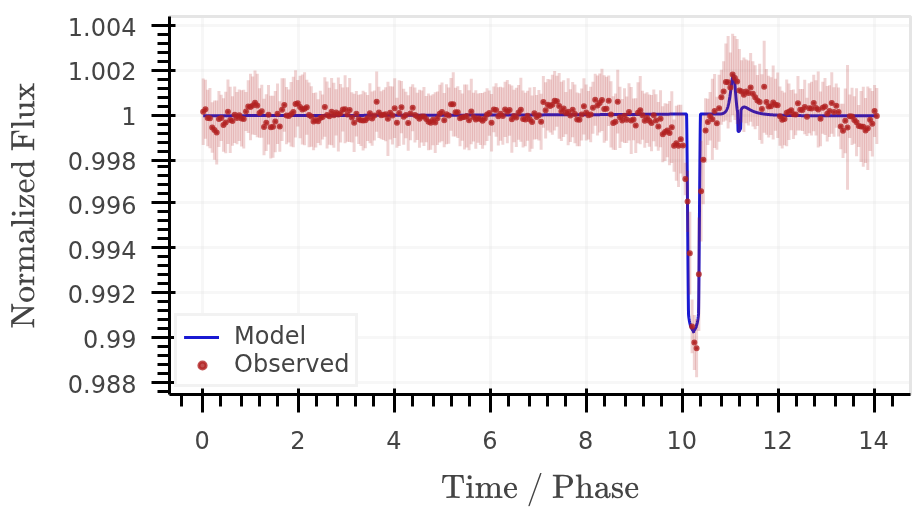}
    \caption{Folded \gls{TESS} light curve of TIC 273777583. This candidate was rejected during manual vetting because the observed variability is not well described by the heartbeat-binary model.}
    \label{fig:weird-signal}
\end{figure}

\begin{figure}
    \centering
    \includegraphics[width=0.9\linewidth]{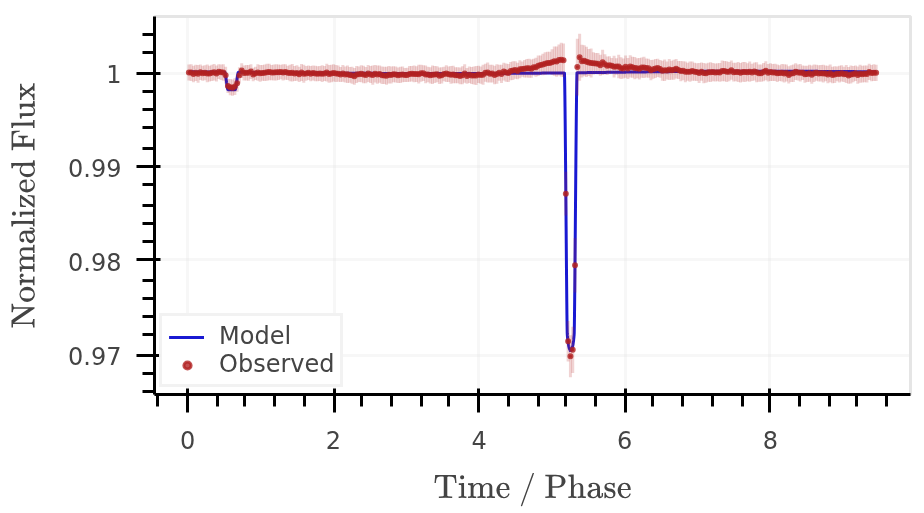}
    \caption{Folded \gls{TESS} light curve of TIC 167201539. This candidate was rejected during manual vetting because the heartbeat-binary model provides a poor fit to the observed signal.}
    \label{fig:straight-across}
\end{figure}

Some of the models produced by the MCMC did not adequately reproduce the observed phase-folded light curves. We identified several types of poor fits. In some cases, the model failed to reproduce one or more dips present in the observed light curve, as shown in Figure~\ref{fig:missing-dip}. In other cases, the model remained within the observational error bars but produced additional features that were not present in the data, as shown in Figure~\ref{fig:weird-signal}. Finally, some models remained within the error bars but consistently followed their boundaries rather than the overall shape of the observed light curve, as shown in Figure~\ref{fig:straight-across}.


\begin{figure}
    \centering
    \includegraphics[width=\linewidth]{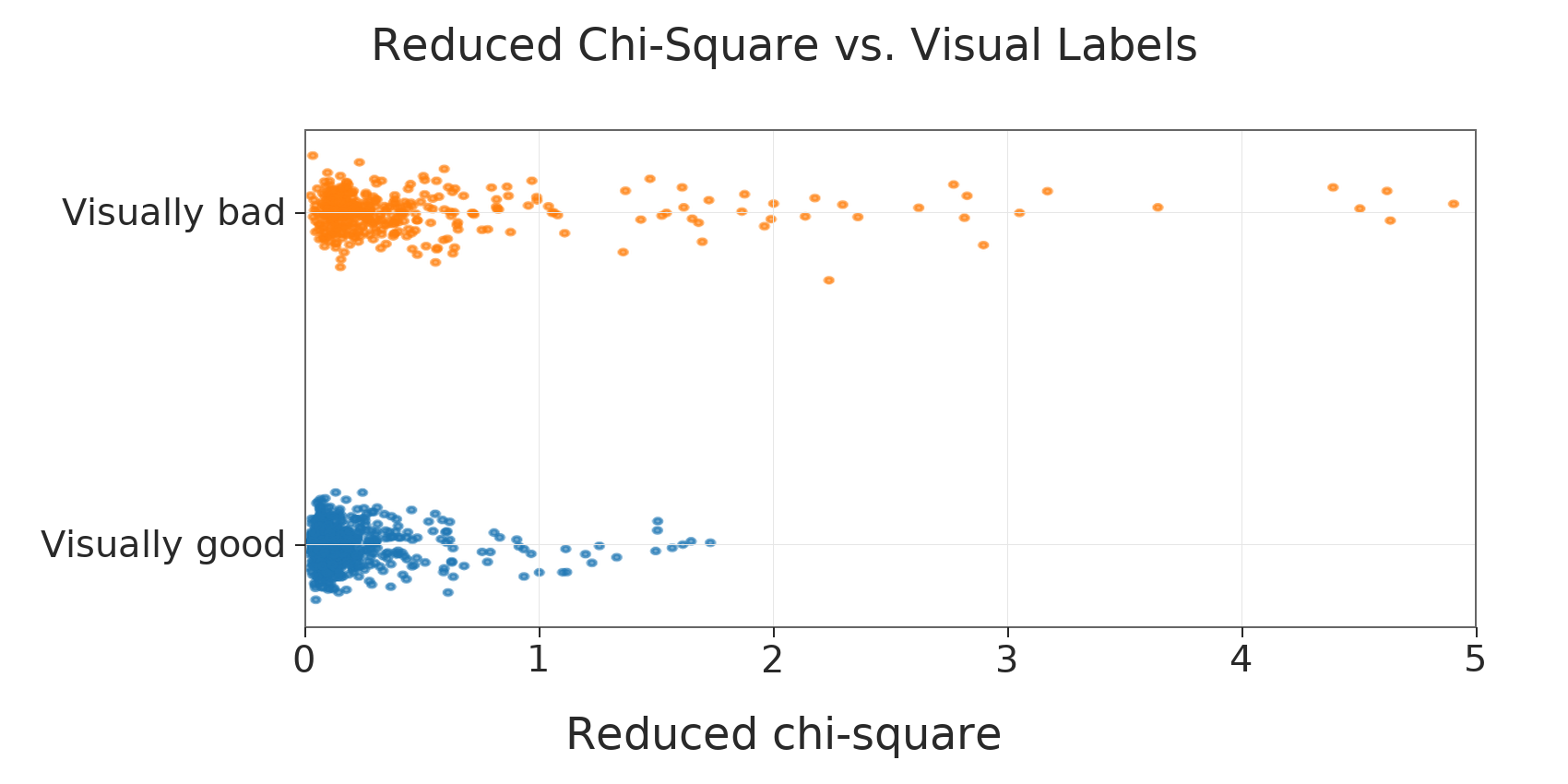}
    \caption{Reduced chi-square values compared with visual model-quality classifications. Each point represents a fitted heartbeat binary model, classified as visually good (blue) or visually bad (orange). Reduced chi-square was calculated from the squared residuals between the observed and modeled light curves, weighted by the observational uncertainties and divided by the degrees of freedom. The overlap between the two groups demonstrates that reduced chi-square alone does not reliably distinguish visually good fits from models exhibiting systematic discrepancies.}
    \label{fig:red-chi}
\end{figure}

\begin{figure}
    \centering
    \includegraphics[width=\linewidth]{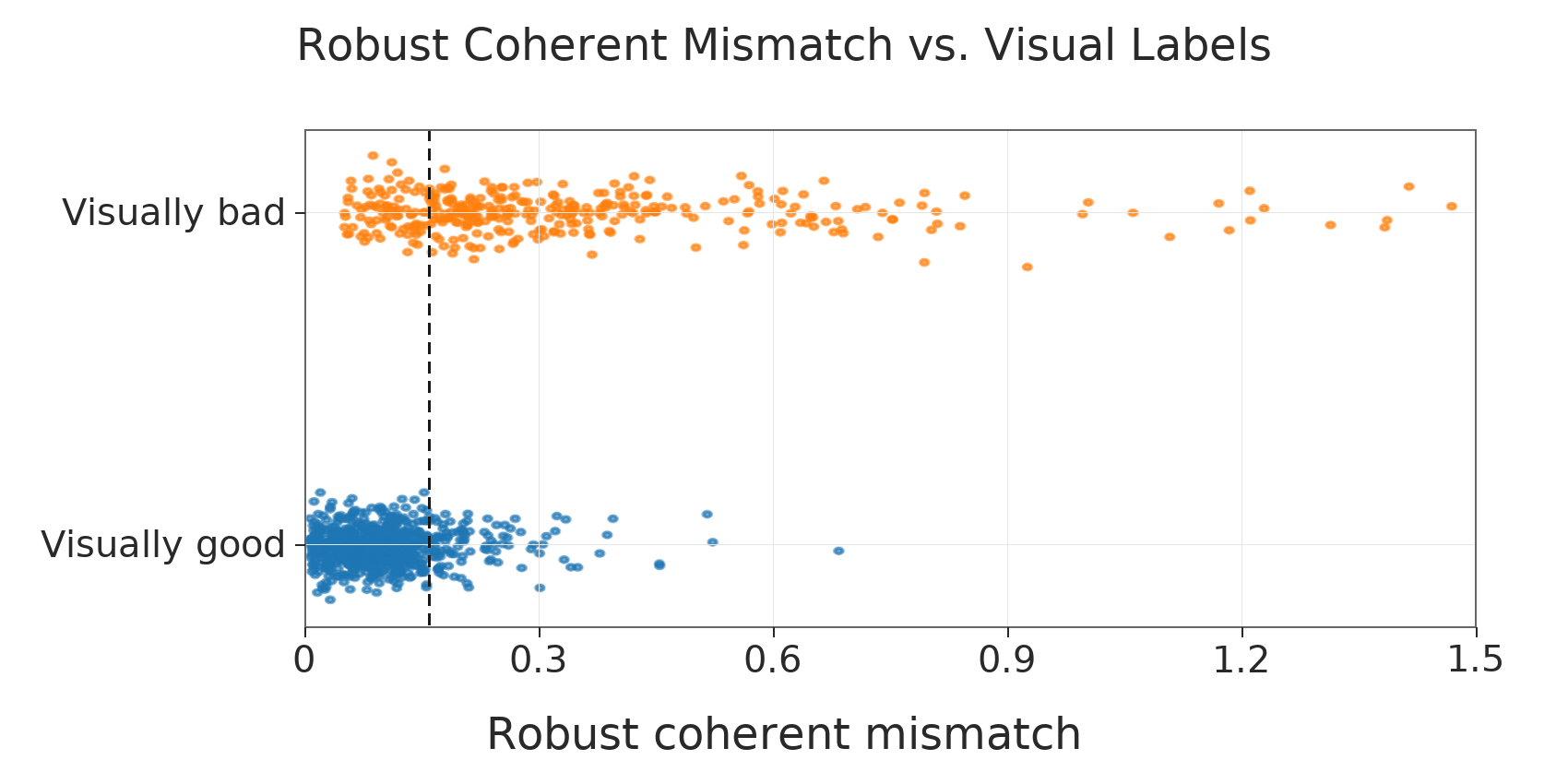}
    \caption{Robust coherent mismatch metric compared with visual model-quality classifications. Each point represents a fitted heartbeat binary model, classified as visually good (blue) or visually bad (orange). The metric is calculated as the 99th percentile of the absolute mean residual within a sliding window spanning 2\% of the orbital phase, normalized by the observed light-curve amplitude (99th minus 1st percentile of the observed flux). The dashed vertical line indicates a reference value of 0.16. Lower values indicate better agreement between the model and observed light curve, while higher values indicate greater systematic deviations.} 
    \label{fig:robust-classification}
\end{figure}

Based on the visual inspection of the fitted light curves, we classified 732 models as good and 371 as bad. Models were classified as good when they adequately reproduced the principal photometric features of the observed light curves, and as bad when they exhibited discrepancies such as those illustrated in the preceding figures. These visual classifications, shown in blue and orange, respectively, were assigned independently of the quantitative goodness-of-fit metrics discussed below. To further evaluate the quality of the fitted models, we initially considered using reduced $\chi^2$ as a measure of goodness of fit. However, we found that this statistic was not always effective at identifying systematic mismatches between the model and observations. In some cases, the model closely followed the observed light curve over most of the orbital phase but failed to reproduce specific features, resulting in localized discrepancies that were not adequately captured by an overall goodness-of-fit statistic, as reflected in the overlap between visually good and bad fits in Figure~\ref{fig:red-chi}. To better identify these discrepancies and provide a reproducible quantitative assessment complementary to visual inspection, we developed a robust coherent mismatch metric that measures systematic differences between the observed and modeled light curves. 
We first calculated the residuals by subtracting the model flux from the observed flux,
\begin{equation}
    r_i = F_i^{\mathrm{obs}} - F_i^{\mathrm{mod}}.
\end{equation}
We then computed the mean residual within a sliding window covering 2\% of the orbital phase, accounting for the periodic nature of the folded light curve. The local mean residual for window $W_j$ is given by
\begin{equation}
    \bar{r}_j = \frac{1}{N_j}\sum_{i \in W_j} r_i,
\end{equation}
where $N_j$ is the number of data points within the window. The robust coherent mismatch metric is then defined as
\begin{equation}
    \mathcal{M} =
    \frac{P_{99}(|\bar{r}|)}
    {P_{99}(F^{\mathrm{obs}})-P_{1}(F^{\mathrm{obs}})},
    \label{eq:robust-mismatch}
\end{equation}
where $P_k$ denotes the $k$th percentile of the corresponding distribution. Lower metric values indicate smaller coherent deviations between the model and observations, while higher values indicate more substantial systematic mismatches, as shown in Figure~\ref{fig:robust-classification}. A reference value of 0.16 was selected by visually inspecting the distribution of mismatch scores for models classified as good or bad. This value was chosen as an approximate boundary between the two groups rather than through a formal optimization procedure. Using this reference, 86.61\% of visually good models (634 out of 732) had mismatch scores below 0.16, while 77.09\% of visually bad models (286 out of 371) had scores above 0.16. Overall, the metric agreed with the visual classifications for 83.41\% of the evaluated models (920 out of 1,100). These results suggest that the metric provides a useful, reproducible quantitative measure of model quality that complements visual inspection and can reduce reliance on subjective judgments, although some overlap remains between the two groups.

    \section{Results}
\label{sec:results}

Our search identified approximately 1,100 potential heartbeat binary candidates in the \gls{TESS} light-curve sample. Following visual inspection and MCMC modeling, 688 systems were retained in the final sample based on the quality of their model fits and a minimum median posterior eccentricity of $e = 0.2$. The fitted parameters provide estimates of the orbital and stellar properties of these systems and allow us to investigate trends within the selected sample. The complete catalog of 688 heartbeat binary candidates, including the fitted stellar and orbital parameters and their central 68.27\% posterior credible intervals where applicable, is provided in machine-readable format as an ancillary file accompanying this article.

\begin{figure*}
    \centering

    \begin{minipage}{0.3\textwidth}
        \centering
        \includegraphics[width=\linewidth]{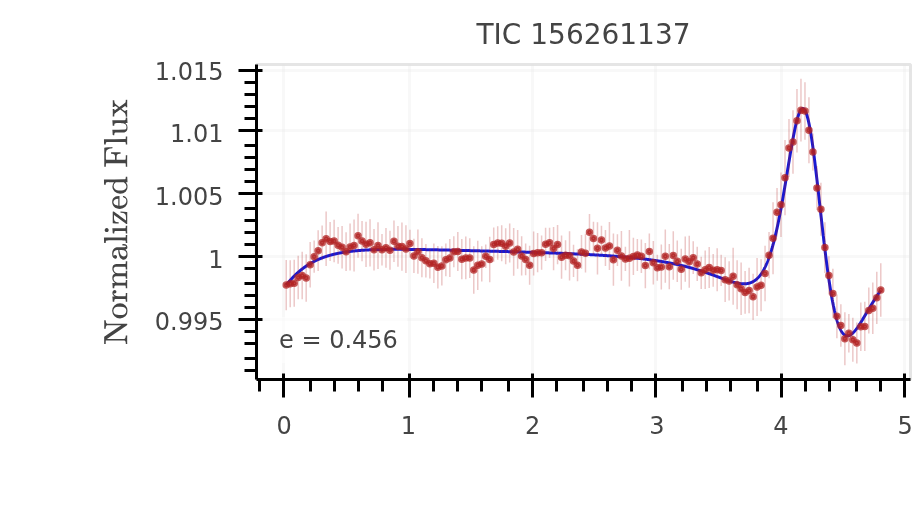}
    \end{minipage}
    \vspace{0.1cm}
    \begin{minipage}{0.3\textwidth}
        \centering
        \includegraphics[width=\linewidth]{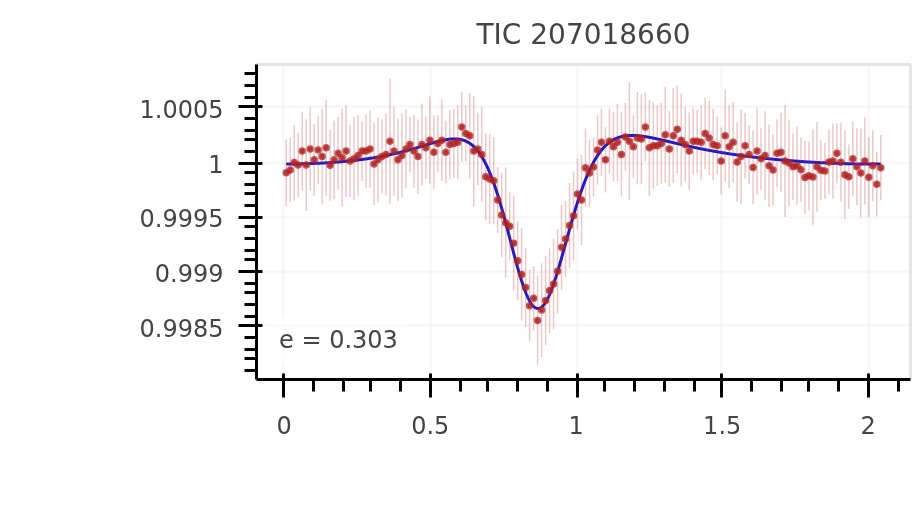}
    \end{minipage}

    \vspace{0.15cm}

    \begin{minipage}{0.3\textwidth}
        \centering
        \includegraphics[width=\linewidth]{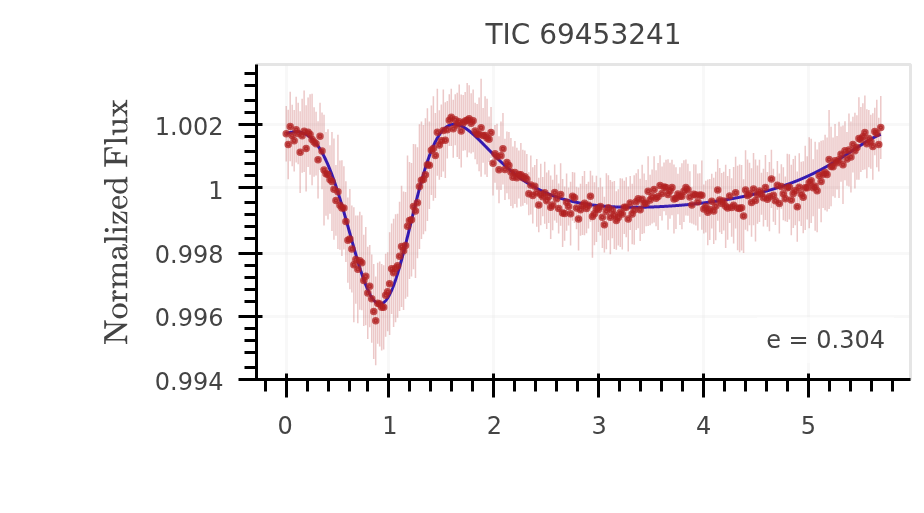}
    \end{minipage}
    \vspace{0.1cm}
    \begin{minipage}{0.3\textwidth}
        \centering
        \includegraphics[width=\linewidth]{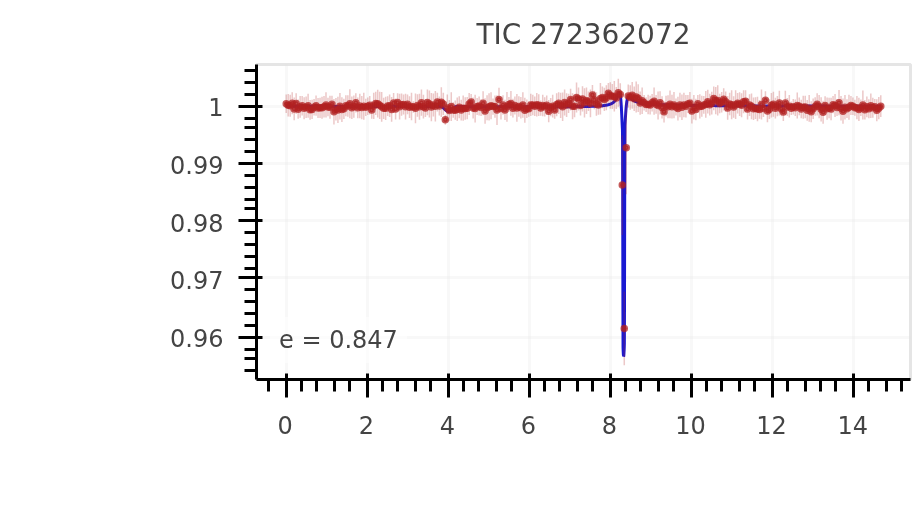}
    \end{minipage}

    \vspace{0.15cm}

    \begin{minipage}{0.3\textwidth}
        \centering
        \includegraphics[width=\linewidth]{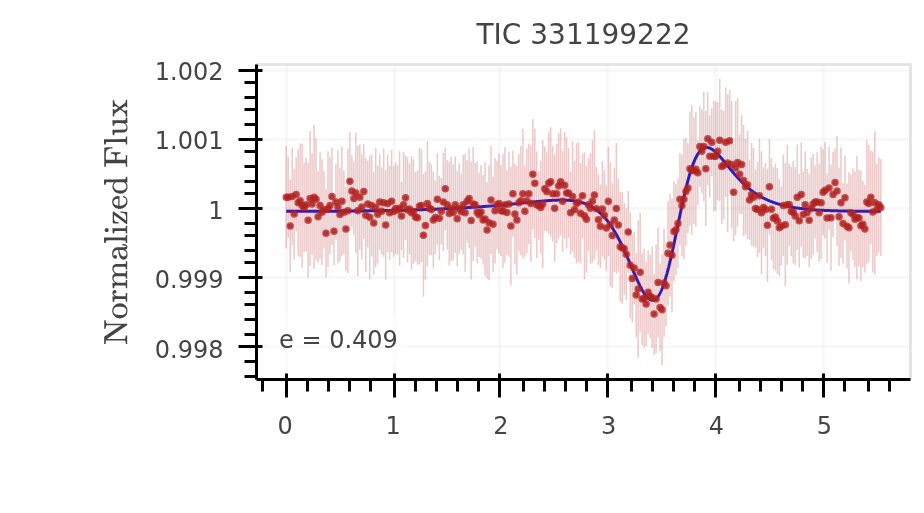}
    \end{minipage}
    \vspace{0.1cm}
    \begin{minipage}{0.3\textwidth}
        \centering
        \includegraphics[width=\linewidth]{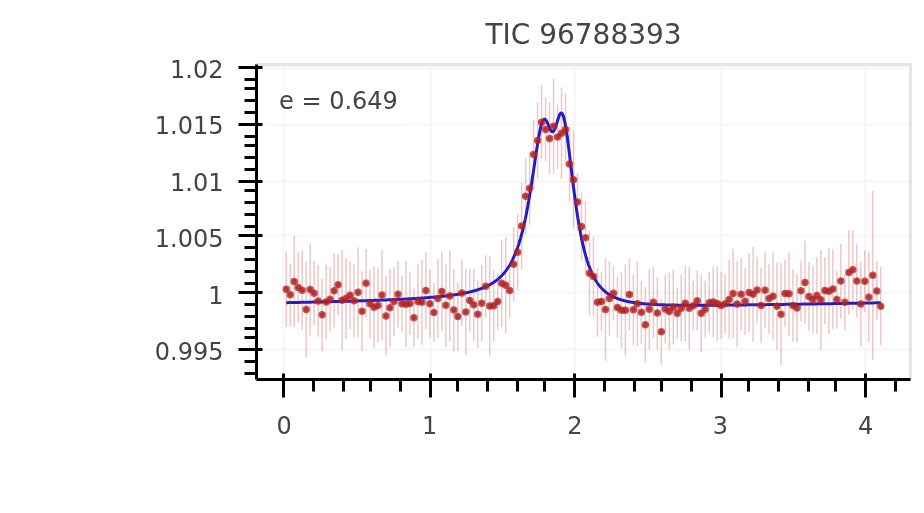}
    \end{minipage}

    \vspace{0.15cm}

    \begin{minipage}{0.3\textwidth}
        \centering
        \includegraphics[width=\linewidth]{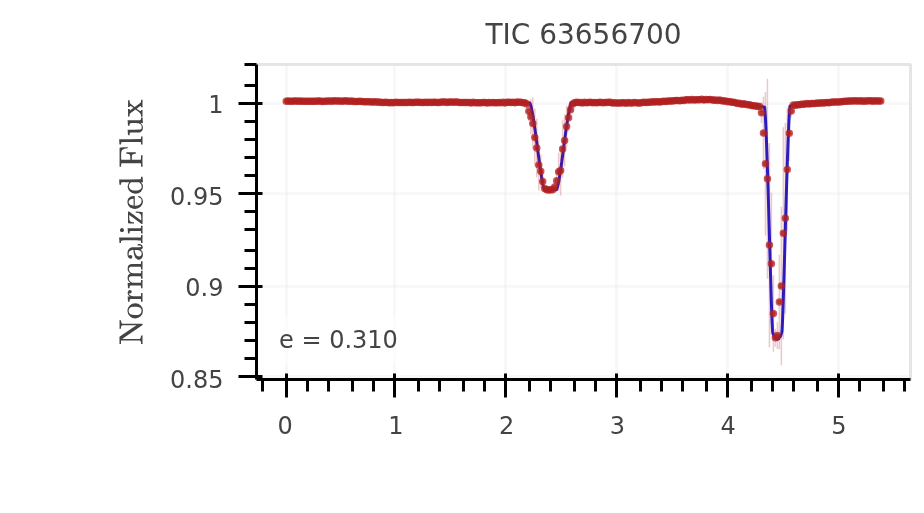}
    \end{minipage}
    \vspace{0.1cm}
    \begin{minipage}{0.3\textwidth}
        \centering
        \includegraphics[width=\linewidth]{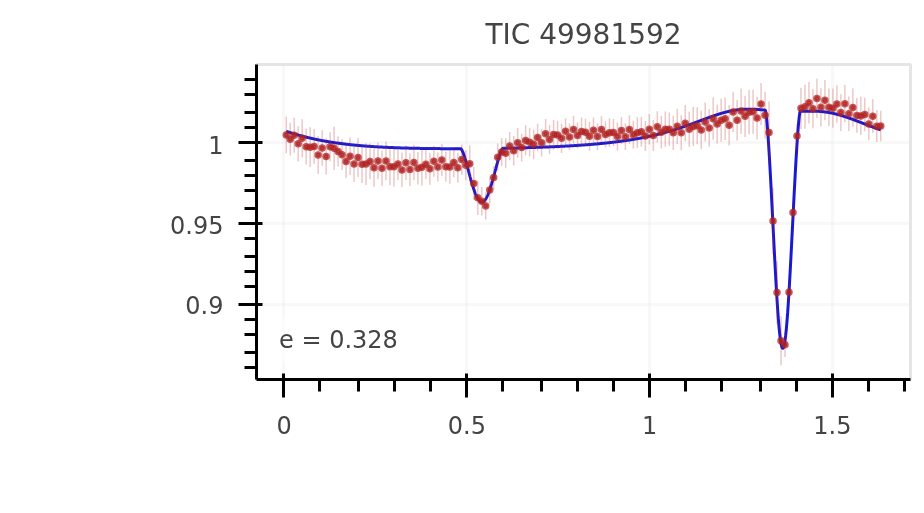}
    \end{minipage}

    \vspace{0.15cm}

    \begin{minipage}{0.3\textwidth}
        \centering
        \includegraphics[width=\linewidth]{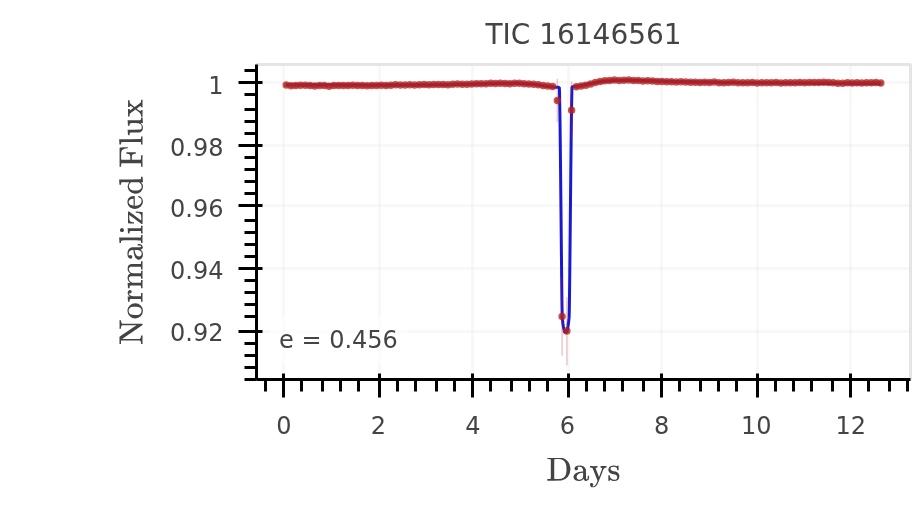}
    \end{minipage}
    \vspace{0.1cm}
    \begin{minipage}{0.3\textwidth}
        \centering
        \includegraphics[width=\linewidth]{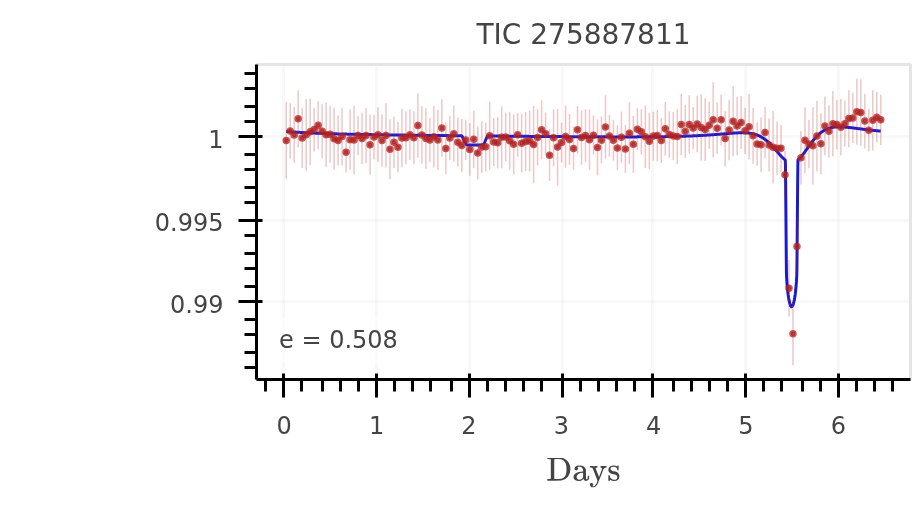}
    \end{minipage}

    \caption{Examples of 10 heartbeat binary candidates and their fitted light-curve models. The observed phase-folded \gls{TESS} light curves are shown together with the corresponding best-fit models. These systems illustrate the range of heartbeat morphologies present in the final sample and the ability of the model to reproduce the observed variability.}
    \label{fig:model-examples}
\end{figure*}

Figure~\ref{fig:model-examples} presents the phase-folded light curves and corresponding highest-likelihood models for 10 randomly selected heartbeat binary candidates from the final sample. The examples illustrate the diversity of observed heartbeat morphologies, including differences in the shapes, amplitudes, and widths of the photometric variations. The fitted models reproduce the principal features of the observed light curves, demonstrating the range of signals that can be characterized using the adopted modeling procedure.

\begin{figure*}
    \centering
    \includegraphics[width=0.9\linewidth]{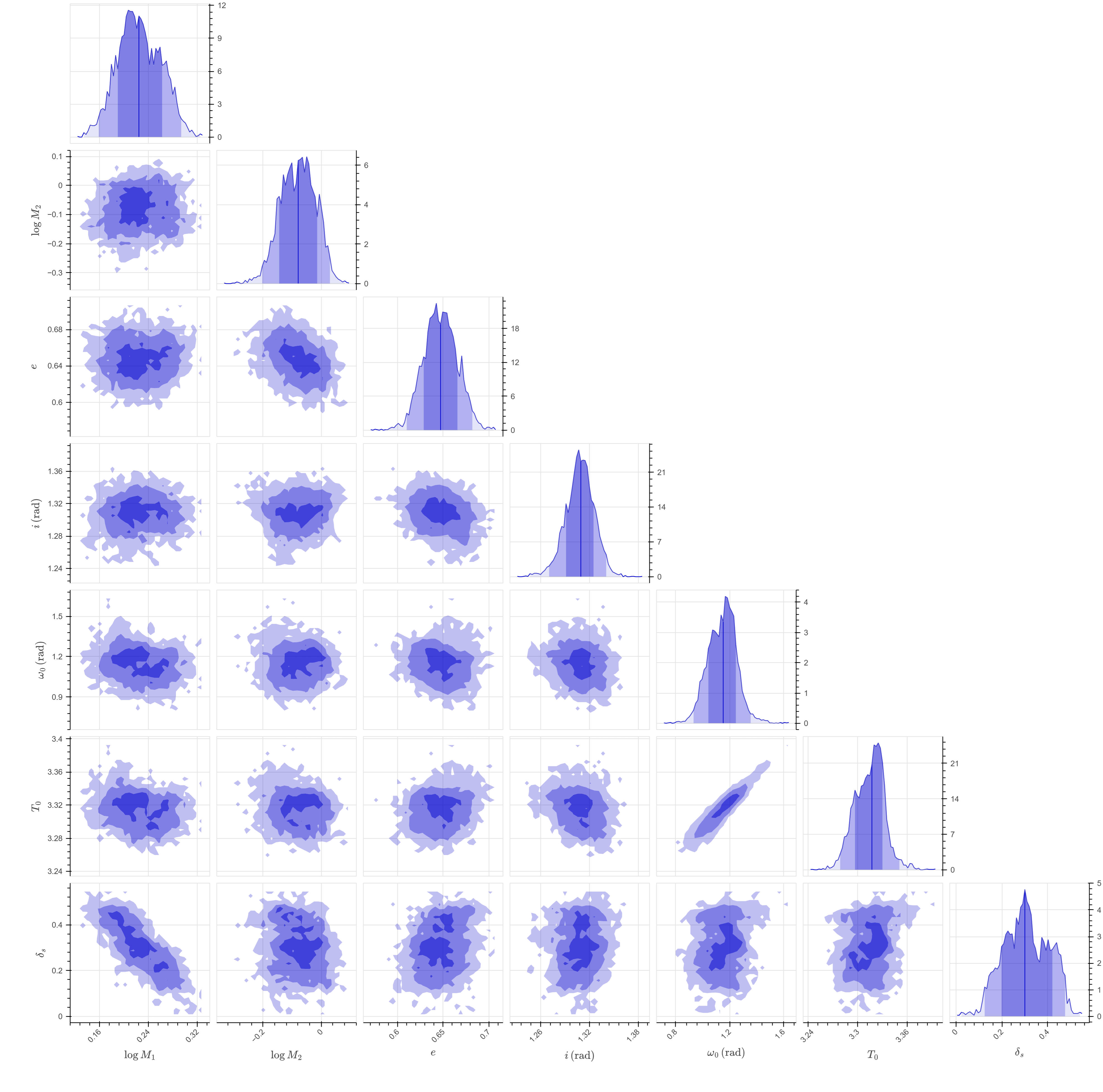}
    \caption{Corner plot showing the posterior distributions for seven selected fitted parameters for TIC 302795956: the logarithms of the primary and secondary stellar masses $\log M_1$ and $\log M_2$, orbital eccentricity $e$, orbital inclination $i$, argument of periastron $\omega_0$, epoch of pericenter passage $T_0$, and blending parameter $\delta_s$.}
    \label{fig:corner-plot}
\end{figure*}

Figure~\ref{fig:corner-plot} shows the marginalized posterior distributions and pairwise correlations of seven selected model parameters for TIC 302795956. The parameters shown are the logarithms of the primary and secondary stellar masses, orbital eccentricity, orbital inclination, argument of periastron, epoch of pericenter passage, and the extrinsic blending parameter $\delta_s$. The diagonal panels show the marginalized posterior distributions, while the off-diagonal panels illustrate correlations and degeneracies between fitted parameters. Most of the marginalized posterior distributions are approximately Gaussian, although the distribution of $\delta_s$ is noticeably non-Gaussian, and many parameter pairs exhibit relatively weak correlations. However, $\omega_0$ and $T_0$ are strongly correlated, reflecting their combined influence on the phase of the modeled photometric features. The primary stellar mass and blending parameter $\delta_s$ are strongly anti-correlated, consistent with a degeneracy between the intrinsic stellar flux contribution and blending within the TESS aperture.

\begin{figure*}
    \centering
    \includegraphics[width=0.7\textwidth]{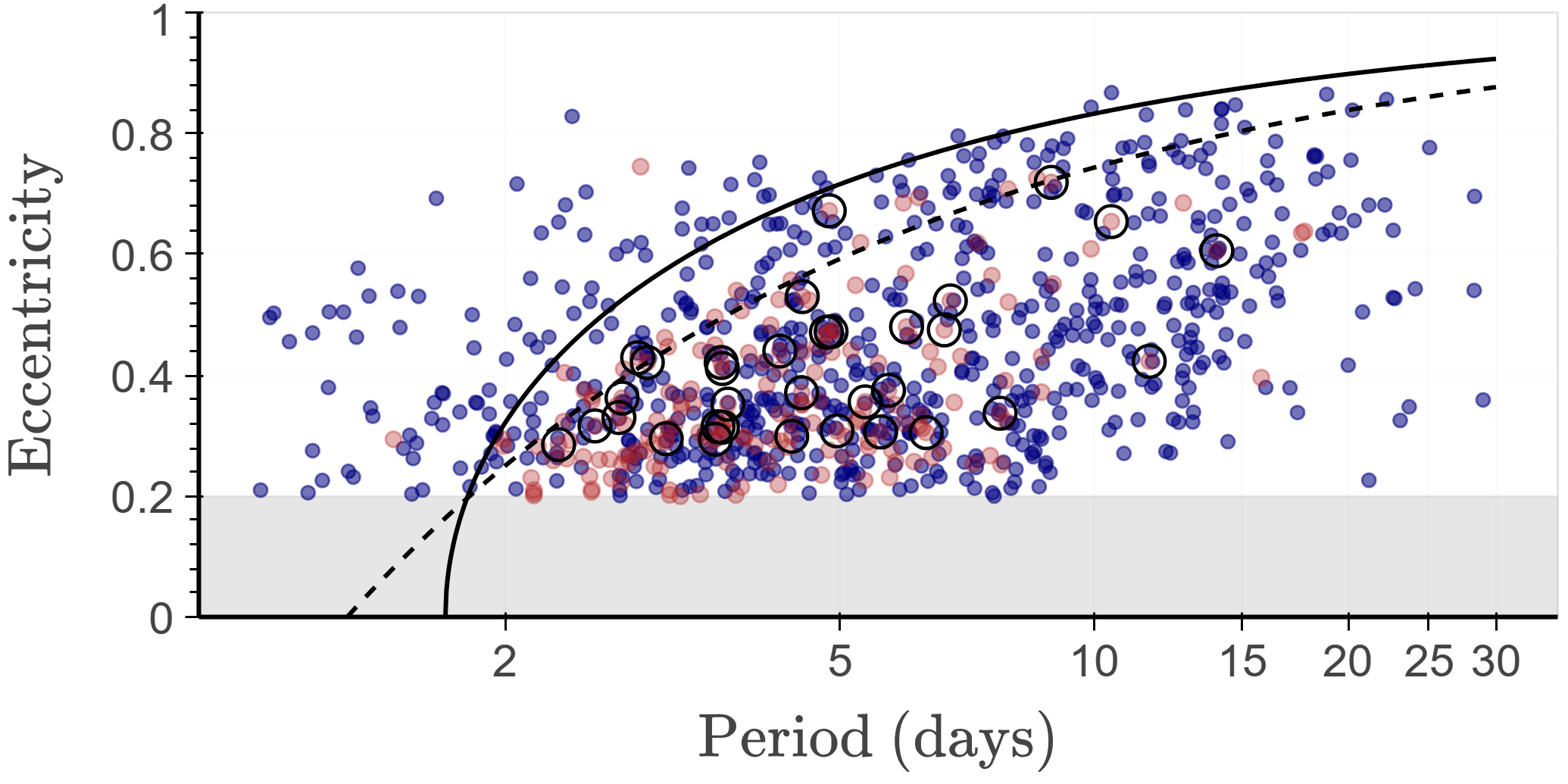}
    \caption{Orbital period--eccentricity distribution of the heartbeat binary candidates identified in this study (blue) compared with the sample of \citet{solanki2025short} (red). The 33 systems common to both samples are outlined in black. The shaded region indicates eccentricities below $e=0.2$, which were excluded from our final sample.}
    \label{fig:period-vs-ecc}
\end{figure*}

Figure~\ref{fig:period-vs-ecc} shows the orbital period--eccentricity distribution of our heartbeat binary candidates compared with the sample of 180 heartbeat binaries identified by \citet{solanki2025short}. The orbital periods correspond to those adopted from the period search and held fixed during MCMC fitting, while the eccentricities are the median values of the MCMC posterior distributions. Our sample is concentrated primarily at shorter orbital periods, with most systems having periods below 10 days, although a smaller number extend beyond 20 days. The comparison sample overlaps substantially with ours in the period--eccentricity plane, while our larger sample extends across a broader range of periods and eccentricities. Nevertheless, only 33 systems are common to both samples, corresponding to approximately 18\% of the systems identified by \citet{solanki2025short}. Of the 688 heartbeat binary candidates identified in this work, 655 are not included in their catalog. Our catalog contains nearly four times as many candidates as the earlier catalog, substantially expanding the population of heartbeat binary candidates identified in TESS data.

The distribution shows a general increase in the upper range of observed eccentricities toward longer orbital periods, although substantial scatter is present. The solid and dashed black curves show the constant orbital angular momentum and constant periastron distance relations, respectively, fitted by \citet{solanki2025short} to the upper eccentricity envelope of their sample. These curves provide empirical reference relations rather than strict physical boundaries. Several candidates in our
sample, particularly at short orbital periods,
lie above these curves.
The gray shaded region indicates eccentricities of $e \leq 0.2$, below the threshold adopted for inclusion in our final sample. 

\begin{figure}
    \centering
    \includegraphics[width=0.9\linewidth]{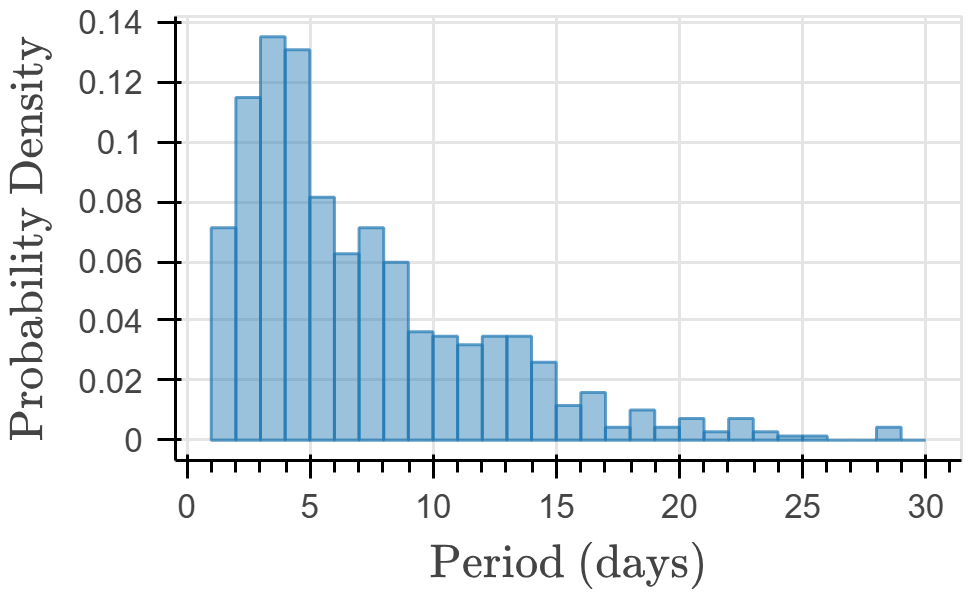}
    \caption{Distribution of the best orbital periods for the heartbeat binary candidates with median posterior eccentricities of $e \geq 0.2$.}
    \label{fig:period-distribution}
\end{figure}

Figure~\ref{fig:period-distribution} shows the distribution of orbital periods for the final heartbeat binary sample. Most candidates have relatively short orbital periods, with the distribution peaking at approximately 3--5 days. The number of systems generally decreases toward longer periods, although a smaller number of candidates have orbital periods exceeding 20 days. As with transiting exoplanets, there is an observational selection bias toward systems with shorter orbital periods, as more orbital cycles can be observed within a given observational baseline. The turnover at $P\lesssim 4$ days may reflect the increasing efficiency of tidal circularization at smaller orbital separations, which would make heartbeat binaries intrinsically less common at such short orbital periods. However, the contribution of selection effects to this turnover cannot be excluded.

\begin{figure}
    \centering
    \includegraphics[width=0.9\linewidth]{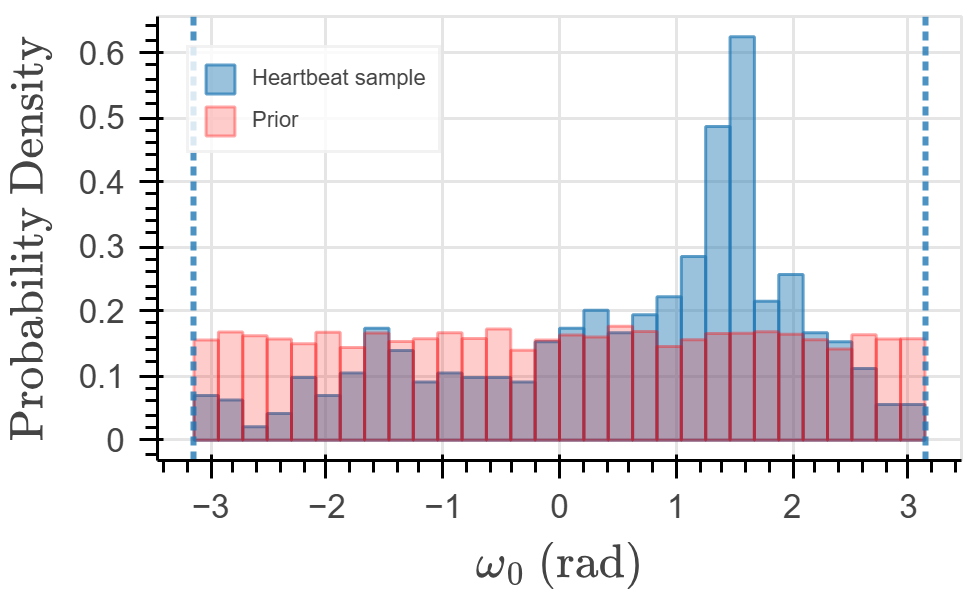}
    \caption{Distribution of the posterior median argument of periastron, $\omega_0$, for the heartbeat binary candidates (blue), compared with the uniform prior distribution adopted during MCMC fitting (red).}
    \label{fig:omega-distribution}
\end{figure}

Figure~\ref{fig:omega-distribution} shows the distribution of the posterior median argument of periastron, $\omega_0$, for the final heartbeat binary sample compared with the uniform prior adopted during MCMC fitting. The distribution exhibits a pronounced peak near $\omega_0 = \pi/2$, with a larger concentration of systems at positive values of $\omega_0$ than at negative values. This differs substantially from the uniform prior distribution. Since the argument of periastron influences the morphology and amplitude of the observed heartbeat signal, the inferred distribution
may reflect selection effects associated with the
sensitivity of our detection and modeling procedures to different
orbital orientations.

\begin{figure}
    \centering
    \includegraphics[width=0.9\linewidth]{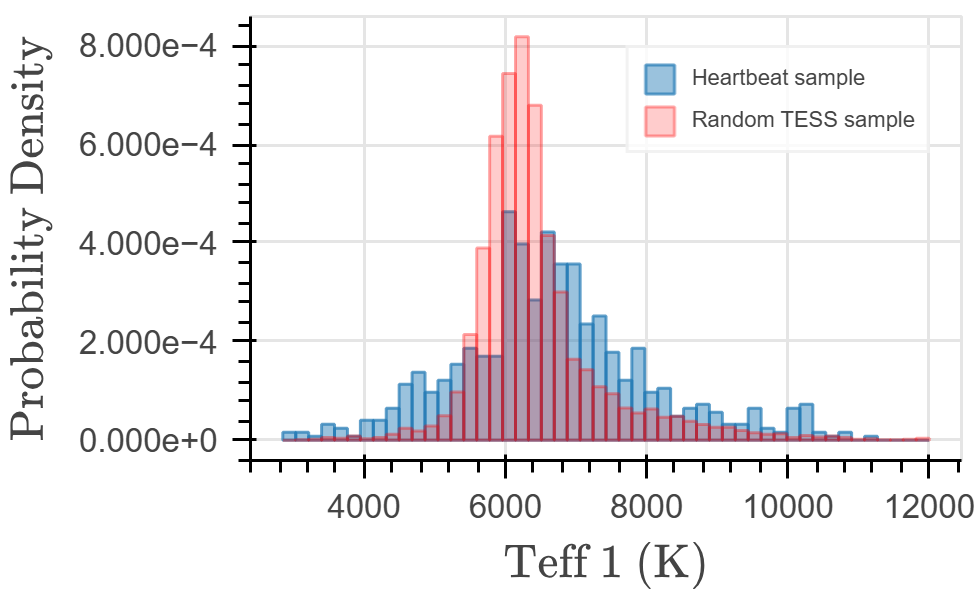}
    \caption{Distribution of the derived effective temperature $T_{eff,1}$ of primary star compared with a random TESS sample.}
    \label{fig:teff1}
\end{figure}
\begin{figure}
    \centering
    \includegraphics[width=0.9\linewidth]{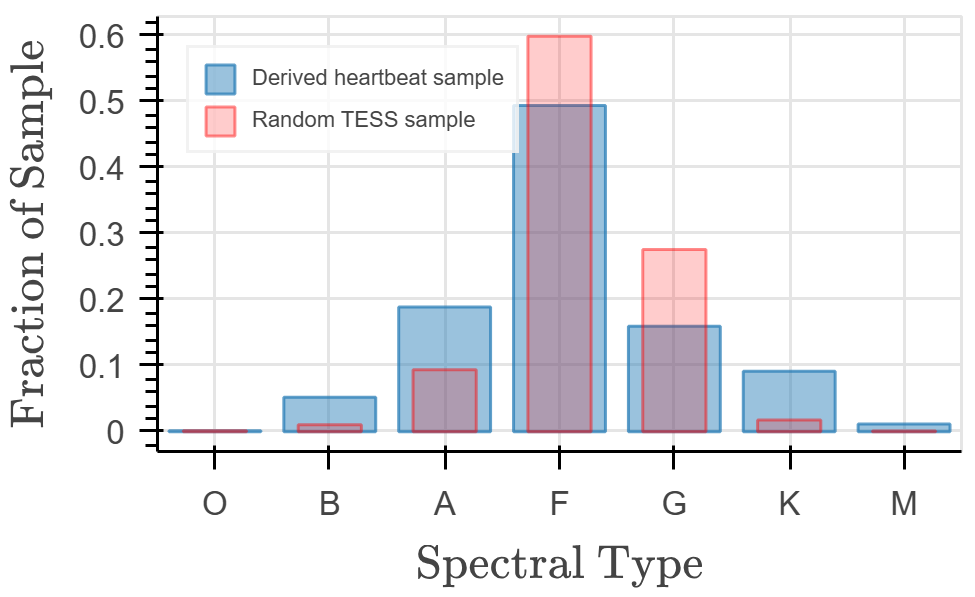}
    \caption{Distribution of the estimated spectral types for the primary star inferred from the derived $T_{eff,1}$ values compared with a random TESS sample.}
    \label{fig:teff1_spectral}
\end{figure}

The effective temperatures and radii of both stellar components were derived from the median posterior model parameters rather than taken directly from the TESS Input Catalog (TIC). For each component, the posterior median stellar mass was used to determine a baseline effective temperature through interpolation of the model's tabulated mass--temperature relation. The fitted temperature rescaling parameter was then applied according to
\begin{equation}
T_{\mathrm{eff}} = 10^{T(M)+0.0224\alpha_T}\, ,
\end{equation}
where $T(M)$ is the base-10 logarithm of the baseline effective temperature in kelvin predicted for a star of mass $M$, $\alpha_T$ is the posterior median temperature rescaling parameter, and $0.0224$ is a fixed coefficient controlling the temperature adjustment in logarithmic space \citep{solanki2025short}.

Similarly, the stellar radii were calculated using the posterior median masses and radius rescaling parameters according to
\begin{equation}
R = 10^{R(M)+rE(M)}\, ,
\end{equation}
where $R(M)$ is the base-10 logarithm of the baseline radius in solar-radius units obtained from the model's tabulated mass--radius relation, $r$ is the posterior median radius rescaling parameter, and $E(M)$ is a mass-dependent scaling function that determines the magnitude of the radius adjustment \citep{solanki2025short}. Both relations were applied independently to the primary and secondary stellar components. The inferred temperatures and radii thus depend on the information provided by the light curves, the adopted mass--temperature and mass--radius relations, and the parameter priors.

Figure~\ref{fig:teff1} compares the model-derived effective temperatures of the primary stars in our heartbeat binary sample (blue) with those of a randomly constructed \gls{TESS} comparison sample (red). To construct the comparison sample, approximately 10,000 TESS-SPOC FFI light curves from Sectors 27--55 were randomly selected from the same light-curve dataset used in our search. 
Targets with valid effective temperature estimates in the TESS Input Catalog (TIC) were then randomly paired to create a reference distribution of hypothetical binary systems. Within each pair, the hotter star was assigned to the $T_{\mathrm{eff},1}$ distribution, while the cooler star was assigned to the $T_{\mathrm{eff},2}$ distribution. This comparison sample provides a reference based on the observed \gls{TESS} stellar population rather than a physically motivated population of binary systems. The primary-star temperatures of the comparison sample are concentrated around 6,500 K, whereas the model-derived temperatures of the heartbeat primaries span a broader range, with a more pronounced high-temperature tail. The heartbeat primary-star distribution also appears shifted toward higher temperatures relative to the comparison sample.

Figure~\ref{fig:teff1_spectral} shows the spectral-type distributions of the primary stars, inferred from their effective temperatures, for both the heartbeat binary sample (blue) and the randomly paired \gls{TESS} comparison sample (red). F-type stars constitute the largest fraction of primary stars in both populations, accounting for approximately 50\% of the heartbeat sample and 60\% of the comparison sample. However, the heartbeat sample contains relatively larger fractions of A-, B-, and K-type stars. This indicates a broader representation of primary-star spectral types in the heartbeat sample, consistent with the differences observed in the effective temperature distributions.

\begin{figure}
    \centering
    \includegraphics[width=0.9\linewidth]{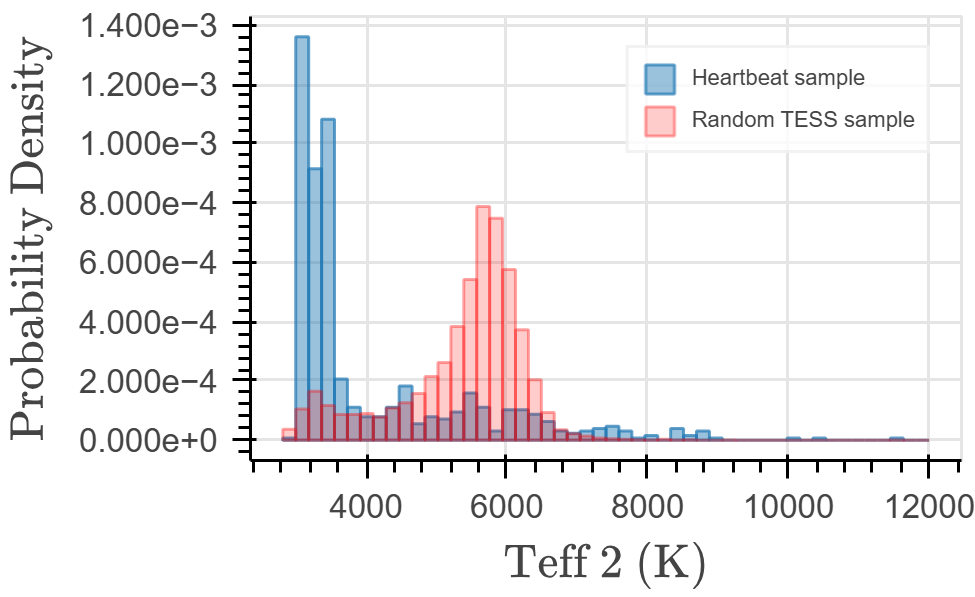}
    \caption{Distribution of the derived effective temperature $T_{eff,2}$ of secondary star compared with a random TESS sample.}
    \label{fig:teff2}
\end{figure}
\begin{figure}
    \centering
    \includegraphics[width=0.9\linewidth]{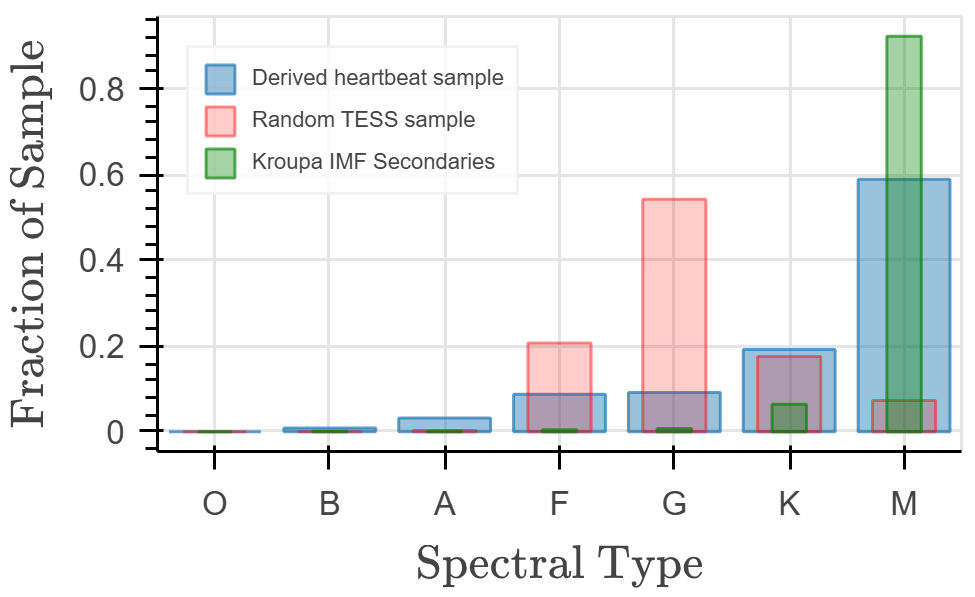}
    \caption{Distribution of the estimated spectral types of the secondary stars inferred from the derived $T_{\mathrm{eff},2}$ values (blue), compared with a randomly paired \gls{TESS} sample (red) and a population drawn from the Kroupa IMF (green).}
    \label{fig:teff2_spectral}
\end{figure}

Figure~\ref{fig:teff2} compares the model-derived effective temperatures of the secondary stars in the heartbeat binary sample with those of the randomly paired \gls{TESS} comparison sample. As described above, the cooler star from each randomly constructed pair was assigned to the $T_{\mathrm{eff},2}$ comparison distribution. Despite this selection, the comparison sample remains concentrated around 5,000--6,000 K, whereas the heartbeat secondary-star distribution exhibits a pronounced peak at substantially lower temperatures, approximately 2,000--3,000 K. The heartbeat sample also extends toward higher effective temperatures, producing a broad distribution with a high-temperature tail.

This difference is further illustrated in Figure~\ref{fig:teff2_spectral}, which presents the corresponding spectral-type distributions inferred from the effective temperatures. More than half of the secondary stars in the heartbeat sample are classified as M-type stars, whereas M-type stars constitute a relatively small fraction of the randomly paired \gls{TESS} comparison sample. The comparison sample is instead dominated by G-type stars, with substantial contributions from F- and K-type stars. These results indicate that the model-derived secondary-star population contains a considerably larger fraction of cool, low-temperature stars than the comparison sample. In Figure~\ref{fig:teff2_spectral} we also plot the spectral-type distribution of a population drawn from the Kroupa IMF \citep{kroupa2001imf}, which is strongly dominated by M-type stars. The relatively large fraction of cool secondary stars inferred in our heartbeat sample is qualitatively consistent with the greater abundance of low-mass stars expected from the IMF and suggests that heartbeat binaries may provide a promising means of identifying faint, low-mass companions through their interactions with brighter stars. The relatively large fraction of cool secondary stars inferred in our heartbeat sample is qualitatively consistent with the greater abundance of low-mass stars expected from the IMF and suggests that heartbeat binaries may provide a promising means of identifying faint, low-mass companions through their interactions with brighter stars.

\begin{figure}
    \centering
    \includegraphics[width=0.9\linewidth]{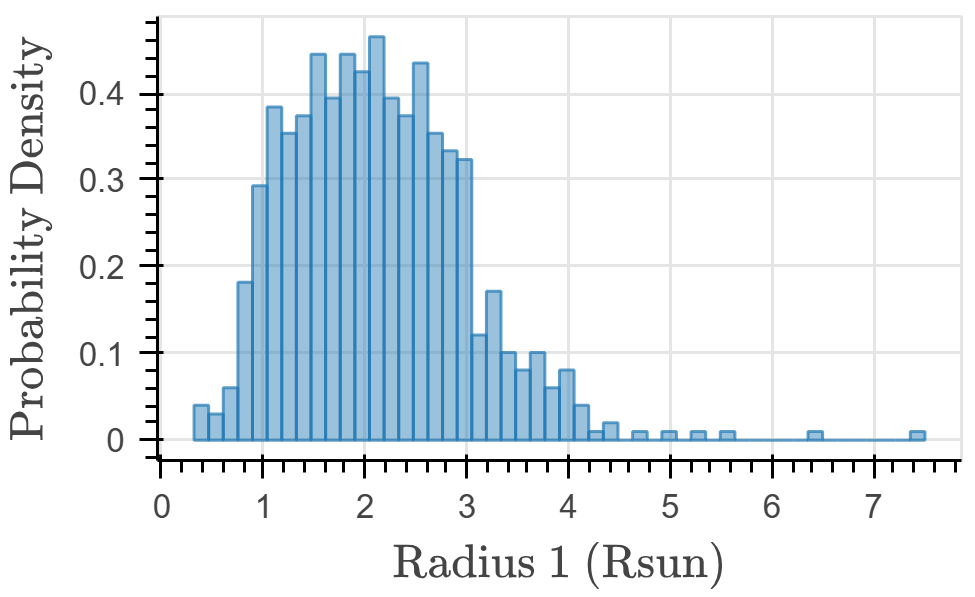}
    \caption{Distribution of the derived radius $R_1$ of the primary stars in our heartbeat binary sample.}
    \label{fig:radius1}
\end{figure}
\begin{figure}
    \centering
    \includegraphics[width=0.9\linewidth]{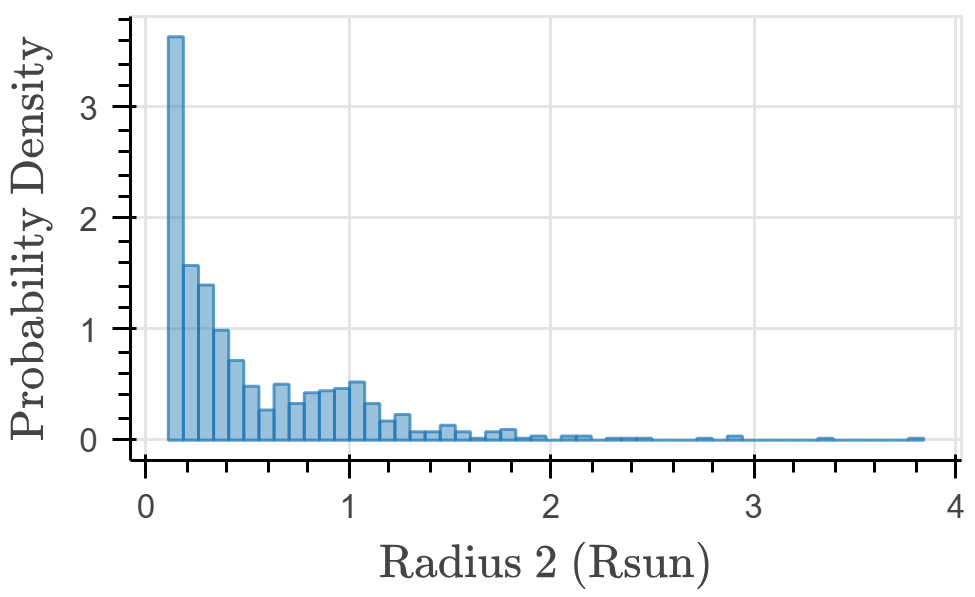}
    \caption{Distribution of the derived radius $R_2$ of the secondary stars in our heartbeat binary sample.}
    \label{fig:radius2}

\end{figure}
The model-derived stellar radius distributions show distinct differences between the primary and secondary components. The primary-star radii, $R_1$, are concentrated between approximately $1.5$ and $2.5\,R_\odot$, with a peak near $2\,R_\odot$ and a tail extending toward larger radii. This range is consistent with the primary-star radii reported by \citet{shporer2016radial}, which range from approximately $1.2$ to $3.3\,R_\odot$. In comparison, \citet{solanki2025short} found a broader range of primary-star radii, with most systems having radii between $1$ and $10\,R_\odot$. This difference is consistent with their sample being predominantly composed of hotter A- and B-type stars. In contrast, the secondary-star radii, $R_2$, in our sample are generally smaller, with a pronounced peak at a few tenths of a solar radius. These results are consistent with the effective-temperature and spectral-type distributions discussed above, in which a substantial fraction of the secondary stars were inferred to be cool, low-mass M-type stars.
    \section{Discussion}
\label{sec:discussion}
The orbital period--eccentricity distribution of our heartbeat binary candidates shows a tendency toward higher eccentricities at longer orbital periods. This trend is qualitatively consistent with the expected effects of tidal interactions in binary systems, which generally become stronger at smaller orbital separations and can lead to orbital circularization over time. The constant-angular-momentum and constant-periastron-distance relations derived from the sample of \citet{solanki2025short} provide useful reference curves for interpreting the observed distribution. Differences in stellar masses and radii between the two samples may also contribute to departures from these reference curves. One possible explanation for systems lying above these reference curves is eccentricity excitation by a third companion, whose gravitational perturbations may sustain high eccentricities despite tidal damping \citep{naoz2016kozai}.

The derived stellar properties provide additional insight into the candidate systems. In particular, the large fraction of secondary components with inferred temperatures corresponding to M-type stars and their relatively small inferred radii suggest that many candidates contain low-mass companions. This is qualitatively consistent with the greater abundance of low-mass stars expected from stellar initial mass functions \citep{kroupa2001imf}. Thus, photometric searches for heartbeat binaries may provide a means of identifying faint, low-mass companions that would otherwise be difficult to detect. However, the mass distribution of these secondary components need not follow the IMF of individual stars, given the mass-ratio distribution of binary systems \citep{moe2017pqs} and the selection effects inherent in our search.

Although the CNN demonstrated strong classification performance on the validation dataset, several limitations must be considered when interpreting the resulting heartbeat binary population. The network was trained to recognize synthetic heartbeat signals injected into real TESS light curves, and its performance therefore depends on how well the injected signals represent the diversity of heartbeat systems in the observed population. Systems with weak heartbeat amplitudes, unusual light-curve morphologies, or substantial observational noise may receive lower confidence scores, even if they are genuine heartbeat binaries. Additionally, heartbeat signals are affected by orbital inclination, eccentricity, and the relative properties of the stellar components, which may influence the probability that a system is identified by the CNN.

In addition to potentially missing genuine heartbeat systems, the CNN may also assign high confidence scores to other sources of astrophysical variability. Eclipsing binaries and planetary transits can produce sharp, periodic changes in brightness, while stellar pulsations and other variable sources may exhibit photometric features that partially resemble heartbeat signals. Instrumental noise and systematic effects may also introduce similar features into the observed light curves. Since the CNN identifies patterns resembling those in its training data rather than determining the physical origin of the variability, high confidence scores do not necessarily correspond to genuine heartbeat binaries. This potential contamination further motivates the manual vetting process, which was used to distinguish plausible heartbeat signals from other forms of variability.

The selection of the 5,000 highest-scoring light curves introduces an additional selection effect, as systems receiving lower scores were excluded from subsequent inspection and modeling. Although this approach substantially reduced the number of light curves requiring manual examination, the resulting sample cannot be assumed to be complete. Furthermore, the CNN confidence scores reflect the network's learned classification function and should not necessarily be interpreted as calibrated probabilities that individual sources are heartbeat binaries. The classification accuracy and AUROC measured on the validation dataset characterize the network's performance on that dataset, but do not independently establish the purity or completeness of the selected TESS candidates.

These limitations are particularly relevant when interpreting the population-level distributions presented in this work. For example, systems with shorter orbital periods may exhibit more repeated heartbeat features within a single TESS observing sector, potentially making them easier for the CNN to recognize than longer-period systems. Similarly, heartbeat signals with larger photometric amplitudes may be more readily distinguished from observational noise. Consequently, the observed distributions of orbital period, eccentricity, and stellar properties may be influenced by the CNN selection process in addition to the underlying astrophysical population (i.e., our survey is subject to selection effects). 

The combined selection effects of the various stages of our search have not been quantified in this work. In addition to the CNN selection effects discussed above, the adopted period-search range and the cutoff on median posterior eccentricity restrict the sample, while systems with ambiguous orbital periods or light curves that are not adequately reproduced by the adopted model may be less likely to be retained. The population distributions presented here should therefore be interpreted as properties of the identified candidate sample rather than unbiased estimates of the intrinsic heartbeat binary population.

An additional limitation arises from the physical modeling of the heartbeat light curves. Although the MCMC procedure provides posterior distributions for the fitted parameters, photometric observations alone may not uniquely constrain all stellar and orbital properties. In particular, degeneracies between stellar masses, radii, orbital inclination, and other model parameters can allow different binary configurations to produce similar light-curve morphologies. The inferred parameters also depend on the adopted priors and the assumptions of the binary-star model. Consequently, the reported posterior estimates should be interpreted within the framework of the adopted model. Additional observations, particularly radial-velocity measurements and spectroscopic constraints on stellar properties, could help distinguish between alternative solutions and improve the determination of individual system parameters.

\section{Conclusion}
\label{sec:conclusion}

We identified 688 heartbeat binary candidates through a search of \gls{TESS} light curves from Sectors 27--55. A convolutional neural network trained on synthetic heartbeat signals injected into observed \gls{TESS} light curves was applied to 4,398,877 TESS-SPOC sector light curves. The 5,000 highest-ranked unique targets were selected for further analysis, substantially reducing the number of light curves requiring manual examination. Orbital periods were estimated using the Box Least Squares algorithm, and the phase-folded light curves were subsequently modeled using a parallel-tempered Markov Chain Monte Carlo (MCMC) approach. The accepted models reproduce a variety of heartbeat morphologies, including narrow photometric features and eclipses. Only systems with acceptable fits and median posterior eccentricities of $e \geq 0.2$ were retained in the final sample.

The inferred orbital parameters reveal several trends within the identified sample. The orbital-period distribution is concentrated at relatively short periods, with a peak around 3--5 days, although some systems have periods exceeding 20 days. The upper eccentricity envelope increases toward longer orbital periods, consistent with the overall trend observed in previously identified heartbeat binary populations. However, a number of candidates, particularly at short orbital periods, lie above the empirical reference curves obtained by \citet{solanki2025short}.

We also used the posterior model parameters to estimate the effective temperatures, spectral types, and radii of both stellar components. The primary-star temperature distribution extends toward higher temperatures than that of the randomly paired \gls{TESS} comparison sample, although F-type stars constitute the largest fraction in both distributions. More pronounced differences are observed for the secondary stars, whose inferred effective temperatures show a strong concentration around 2,000--3,000 K. More than half of the secondary stars have inferred temperatures corresponding to M-type stars, suggesting that many of the identified systems contain cool, low-mass companions. The primary-star radii are concentrated around $1.5$--$2.5\,R_\odot$, while the inferred radii of the secondary stars are generally much smaller, often below $1\,R_\odot$.

Of the 688 heartbeat binary candidates identified in this study, 655 are not included in the catalog of \citet{solanki2025short}, substantially expanding the sample of heartbeat binary candidates identified in \gls{TESS} data. Although the inferred population distributions are affected by selection effects and uncertainties in the light-curve modeling, the catalog provides a useful foundation for further characterization of individual systems and future population studies.

\appendix

\section{Heartbeat Binary Model Parameters}
\label{app:model-parameters}

{
\scriptsize
\setlength{\tabcolsep}{2.5pt}
\begin{longtable}{@{}lcccccc@{}}
\caption{Orbital and stellar parameters of the first 50 accepted heartbeat-binary candidates with $e\geq 0.2$. The reported uncertainties are central 68.27\% posterior credible intervals. The orbital period was held fixed during MCMC fitting. The complete catalog of 688 candidates is available in machine-readable format as an ancillary file accompanying this article.}\label{tab:heartbeat_catalog}\\
\toprule
TIC ID & $P$ (d) & $M_1$ ($M_\odot$) & $M_2$ ($M_\odot$) & $e$ & $i$ ($^\circ$) & $\omega$ ($^\circ$) \\
\midrule
\endfirsthead
\toprule
TIC ID & $P$ (d) & $M_1$ ($M_\odot$) & $M_2$ ($M_\odot$) & $e$ & $i$ ($^\circ$) & $\omega$ ($^\circ$) \\
\midrule
\endhead
1676859 & $8.201$ & $1.752^{+0.165}_{-0.158}$ & $0.319^{+0.078}_{-0.067}$ & $0.508^{+0.016}_{-0.017}$ & $53.3^{+2.4}_{-2.2}$ & $87.0^{+4.2}_{-3.5}$ \\
2020964 & $2.937$ & $2.093^{+0.126}_{-0.156}$ & $0.929^{+0.088}_{-0.074}$ & $0.598^{+0.028}_{-0.038}$ & $88.8^{+0.7}_{-0.7}$ & $-107.3^{+3.2}_{-4.0}$ \\
3550123 & $3.819$ & $1.194^{+0.313}_{-0.065}$ & $0.182^{+0.171}_{-0.138}$ & $0.534^{+0.058}_{-0.371}$ & $65.3^{+9.0}_{-12.3}$ & $96.7^{+7.9}_{-178.2}$ \\
3644236 & $1.771$ & $0.941^{+0.034}_{-0.030}$ & $0.048^{+0.012}_{-0.009}$ & $0.246^{+0.063}_{-0.075}$ & $86.7^{+1.6}_{-1.5}$ & $89.3^{+3.0}_{-3.2}$ \\
4559468 & $21.201$ & $3.632^{+0.738}_{-0.392}$ & $0.356^{+0.060}_{-0.065}$ & $0.681^{+0.032}_{-0.031}$ & $51.0^{+7.5}_{-3.7}$ & $28.4^{+5.2}_{-5.7}$ \\
5162698 & $2.955$ & $1.218^{+0.048}_{-0.090}$ & $0.038^{+0.006}_{-0.004}$ & $0.310^{+0.033}_{-0.036}$ & $81.2^{+3.8}_{-1.8}$ & $79.8^{+7.4}_{-6.5}$ \\
5809203 & $12.610$ & $1.678^{+0.053}_{-0.062}$ & $0.691^{+0.042}_{-0.033}$ & $0.457^{+0.006}_{-0.006}$ & $78.7^{+0.3}_{-0.4}$ & $61.0^{+1.2}_{-1.8}$ \\
6082223 & $4.063$ & $1.010^{+0.041}_{-0.030}$ & $0.937^{+0.042}_{-0.075}$ & $0.359^{+0.001}_{-0.001}$ & $83.5^{+1.1}_{-0.6}$ & $-7.9^{+181.2}_{-1.8}$ \\
6101463 & $5.527$ & $2.130^{+0.066}_{-0.068}$ & $1.591^{+0.088}_{-0.074}$ & $0.363^{+0.014}_{-0.014}$ & $85.0^{+0.4}_{-0.3}$ & $79.6^{+0.5}_{-0.6}$ \\
6276559 & $5.437$ & $1.406^{+0.061}_{-0.124}$ & $0.739^{+0.095}_{-0.097}$ & $0.441^{+0.040}_{-0.027}$ & $76.9^{+0.6}_{-0.9}$ & $66.9^{+2.8}_{-3.9}$ \\
6997783 & $14.495$ & $1.540^{+0.185}_{-0.103}$ & $0.690^{+0.129}_{-0.107}$ & $0.683^{+0.020}_{-0.041}$ & $82.4^{+0.6}_{-0.8}$ & $13.8^{+12.5}_{-5.5}$ \\
7020254 & $20.272$ & $0.648^{+0.027}_{-0.021}$ & $0.168^{+0.015}_{-0.022}$ & $0.838^{+0.047}_{-0.157}$ & $86.7^{+1.7}_{-5.4}$ & $88.1^{+8.6}_{-23.5}$ \\
7696907 & $28.295$ & $0.948^{+0.047}_{-0.042}$ & $0.750^{+0.032}_{-0.067}$ & $0.696^{+0.033}_{-0.067}$ & $85.8^{+0.9}_{-0.6}$ & $-50.8^{+9.1}_{-12.6}$ \\
7827348 & $3.497$ & $1.169^{+0.052}_{-0.047}$ & $0.351^{+0.146}_{-0.091}$ & $0.439^{+0.021}_{-0.021}$ & $67.1^{+3.8}_{-3.8}$ & $86.1^{+4.2}_{-4.4}$ \\
9727392 & $4.535$ & $1.075^{+0.060}_{-0.069}$ & $0.141^{+0.013}_{-0.013}$ & $0.627^{+0.015}_{-0.014}$ & $86.5^{+1.4}_{-0.9}$ & $89.8^{+1.2}_{-1.3}$ \\
10675309 & $1.769$ & $1.494^{+0.077}_{-0.057}$ & $0.038^{+0.005}_{-0.004}$ & $0.339^{+0.027}_{-0.029}$ & $84.4^{+2.7}_{-2.9}$ & $88.9^{+8.2}_{-9.1}$ \\
12339915 & $6.273$ & $1.318^{+0.041}_{-0.051}$ & $0.275^{+0.034}_{-0.028}$ & $0.338^{+0.006}_{-0.007}$ & $67.0^{+3.3}_{-2.8}$ & $-168.8^{+1.8}_{-1.9}$ \\
12854388 & $3.123$ & $1.467^{+0.296}_{-0.331}$ & $0.124^{+0.091}_{-0.057}$ & $0.311^{+0.089}_{-0.135}$ & $62.7^{+8.4}_{-12.5}$ & $46.4^{+34.0}_{-63.4}$ \\
13023378 & $6.067$ & $0.838^{+0.049}_{-0.030}$ & $0.066^{+0.028}_{-0.027}$ & $0.304^{+0.386}_{-0.282}$ & $87.0^{+1.5}_{-11.3}$ & $37.5^{+101.6}_{-7.4}$ \\
13974582 & $4.860$ & $1.983^{+0.125}_{-0.099}$ & $0.932^{+0.091}_{-0.089}$ & $0.339^{+0.017}_{-0.015}$ & $81.6^{+0.9}_{-0.7}$ & $49.7^{+2.6}_{-2.6}$ \\
16146561 & $15.104$ & $1.832^{+0.045}_{-0.052}$ & $1.242^{+0.077}_{-0.134}$ & $0.456^{+0.014}_{-0.016}$ & $88.4^{+1.2}_{-2.1}$ & $89.7^{+3.1}_{-1.4}$ \\
16788147 & $6.160$ & $1.452^{+0.084}_{-0.078}$ & $0.392^{+0.102}_{-0.077}$ & $0.490^{+0.016}_{-0.015}$ & $49.0^{+1.6}_{-1.4}$ & $97.1^{+3.7}_{-3.5}$ \\
16791467 & $8.474$ & $1.259^{+0.055}_{-0.061}$ & $0.172^{+0.025}_{-0.022}$ & $0.265^{+0.012}_{-0.012}$ & $47.7^{+1.8}_{-1.7}$ & $-31.9^{+3.1}_{-3.3}$ \\
17491889 & $10.762$ & $1.192^{+0.189}_{-0.059}$ & $0.097^{+0.142}_{-0.025}$ & $0.554^{+0.034}_{-0.032}$ & $73.9^{+3.6}_{-4.8}$ & $83.5^{+19.3}_{-27.1}$ \\
18891071 & $4.715$ & $1.765^{+0.101}_{-0.085}$ & $1.191^{+0.096}_{-0.087}$ & $0.370^{+0.030}_{-0.017}$ & $84.4^{+0.8}_{-0.7}$ & $126.4^{+2.2}_{-3.6}$ \\
18915304 & $22.248$ & $1.490^{+0.117}_{-0.484}$ & $0.118^{+0.061}_{-0.035}$ & $0.856^{+0.039}_{-0.509}$ & $79.3^{+5.6}_{-26.7}$ & $105.1^{+30.4}_{-145.1}$ \\
19971063 & $13.973$ & $1.637^{+0.048}_{-0.337}$ & $1.586^{+0.065}_{-0.492}$ & $0.603^{+0.001}_{-0.001}$ & $87.3^{+1.0}_{-0.7}$ & $-0.3^{+0.9}_{-178.5}$ \\
21295315 & $11.603$ & $0.863^{+0.019}_{-0.020}$ & $0.650^{+0.034}_{-0.012}$ & $0.667^{+0.032}_{-0.045}$ & $87.7^{+0.1}_{-0.1}$ & $-89.8^{+0.1}_{-0.2}$ \\
21649340 & $8.393$ & $1.390^{+0.041}_{-0.041}$ & $0.169^{+0.017}_{-0.017}$ & $0.363^{+0.013}_{-0.013}$ & $75.9^{+2.4}_{-3.3}$ & $-175.1^{+353.0}_{-3.1}$ \\
22174712 & $3.285$ & $1.343^{+0.088}_{-0.107}$ & $0.179^{+0.049}_{-0.039}$ & $0.520^{+0.025}_{-0.026}$ & $38.1^{+2.1}_{-1.9}$ & $12.4^{+5.8}_{-7.1}$ \\
22254438 & $2.305$ & $1.892^{+0.204}_{-0.200}$ & $0.108^{+0.029}_{-0.028}$ & $0.225^{+0.020}_{-0.018}$ & $45.7^{+2.7}_{-2.4}$ & $65.3^{+5.5}_{-5.3}$ \\
22590328 & $2.737$ & $1.166^{+0.033}_{-0.035}$ & $0.360^{+0.030}_{-0.023}$ & $0.210^{+0.010}_{-0.010}$ & $80.4^{+0.3}_{-0.3}$ & $104.8^{+2.0}_{-2.2}$ \\
22601898 & $8.454$ & $1.561^{+0.066}_{-0.069}$ & $0.399^{+0.048}_{-0.048}$ & $0.472^{+0.011}_{-0.011}$ & $62.0^{+6.1}_{-5.7}$ & $-82.0^{+2.6}_{-2.8}$ \\
22767596 & $2.557$ & $0.866^{+0.027}_{-0.022}$ & $0.371^{+0.042}_{-0.032}$ & $0.381^{+0.012}_{-0.013}$ & $73.9^{+1.2}_{-1.6}$ & $77.4^{+1.9}_{-2.1}$ \\
24776689 & $1.856$ & $0.887^{+0.133}_{-0.125}$ & $0.225^{+0.169}_{-0.132}$ & $0.354^{+0.057}_{-0.054}$ & $80.0^{+1.4}_{-1.4}$ & $39.7^{+9.3}_{-25.6}$ \\
25944308 & $3.947$ & $1.368^{+0.061}_{-0.050}$ & $0.160^{+0.019}_{-0.021}$ & $0.341^{+0.013}_{-0.013}$ & $64.6^{+5.5}_{-4.8}$ & $-63.2^{+3.0}_{-3.2}$ \\
26017005 & $4.098$ & $0.747^{+0.044}_{-0.030}$ & $0.056^{+0.018}_{-0.013}$ & $0.586^{+0.124}_{-0.218}$ & $85.3^{+0.8}_{-1.9}$ & $81.2^{+19.4}_{-22.1}$ \\
26017840 & $1.386$ & $0.369^{+0.041}_{-0.283}$ & $0.192^{+0.026}_{-0.120}$ & $0.346^{+0.464}_{-0.030}$ & $82.3^{+7.3}_{-0.6}$ & $-139.1^{+77.9}_{-5.0}$ \\
26238862 & $12.798$ & $1.475^{+0.083}_{-0.065}$ & $0.514^{+0.096}_{-0.111}$ & $0.593^{+0.021}_{-0.022}$ & $68.8^{+3.6}_{-5.4}$ & $110.4^{+8.3}_{-8.3}$ \\
26466275 & $3.451$ & $1.554^{+0.087}_{-0.076}$ & $1.110^{+0.080}_{-0.069}$ & $0.304^{+0.008}_{-0.007}$ & $76.2^{+0.6}_{-0.5}$ & $154.4^{+2.3}_{-2.3}$ \\
27687462 & $5.800$ & $1.298^{+0.057}_{-0.073}$ & $0.137^{+0.021}_{-0.023}$ & $0.393^{+0.015}_{-0.014}$ & $59.6^{+3.5}_{-3.0}$ & $143.3^{+4.3}_{-3.8}$ \\
28126316 & $4.898$ & $1.380^{+0.066}_{-0.124}$ & $0.049^{+0.015}_{-0.010}$ & $0.654^{+0.012}_{-0.012}$ & $83.6^{+3.4}_{-3.7}$ & $93.0^{+3.6}_{-4.1}$ \\
28830523 & $3.216$ & $1.223^{+0.053}_{-0.056}$ & $0.046^{+0.013}_{-0.009}$ & $0.265^{+0.077}_{-0.047}$ & $88.7^{+0.6}_{-0.5}$ & $-95.2^{+7.8}_{-7.3}$ \\
29067936 & $25.023$ & $1.088^{+0.016}_{-0.178}$ & $0.087^{+0.065}_{-0.048}$ & $0.776^{+0.012}_{-0.483}$ & $89.2^{+0.6}_{-1.1}$ & $-134.0^{+205.4}_{-1.1}$ \\
29119552 & $7.627$ & $1.178^{+0.067}_{-0.051}$ & $0.054^{+0.013}_{-0.011}$ & $0.785^{+0.016}_{-0.020}$ & $85.1^{+1.7}_{-3.5}$ & $83.4^{+10.8}_{-6.9}$ \\
29848696 & $2.537$ & $1.991^{+0.131}_{-0.160}$ & $0.051^{+0.019}_{-0.011}$ & $0.544^{+0.041}_{-0.053}$ & $82.4^{+4.2}_{-5.6}$ & $95.5^{+14.3}_{-12.6}$ \\
33159045 & $7.569$ & $1.439^{+0.079}_{-0.101}$ & $0.362^{+0.040}_{-0.031}$ & $0.713^{+0.019}_{-0.022}$ & $85.5^{+1.6}_{-2.9}$ & $77.9^{+6.9}_{-2.7}$ \\
33241809 & $3.651$ & $2.826^{+0.316}_{-0.223}$ & $0.395^{+0.090}_{-0.064}$ & $0.208^{+0.039}_{-0.036}$ & $56.4^{+7.4}_{-6.5}$ & $136.0^{+13.7}_{-17.7}$ \\
33521996 & $3.323$ & $0.351^{+0.065}_{-0.028}$ & $0.110^{+0.049}_{-0.028}$ & $0.503^{+0.057}_{-0.061}$ & $86.8^{+0.4}_{-0.6}$ & $111.3^{+4.7}_{-236.8}$ \\
34371411 & $3.883$ & $1.177^{+0.050}_{-0.052}$ & $0.049^{+0.021}_{-0.011}$ & $0.340^{+0.297}_{-0.149}$ & $88.7^{+0.7}_{-0.8}$ & $-75.8^{+43.3}_{-23.1}$ \\
\bottomrule
\end{longtable}
}

    \bibliography{bibliography}{}
    \bibliographystyle{aasjournalv7.1}

\end{document}